\documentclass[11pt,a4paper]{article}
\pdfoutput=1 
\usepackage{jheppub}
\usepackage[T1]{fontenc} 
\usepackage{hyperref}
\usepackage{amsmath}
\usepackage{amsfonts}
\usepackage{mathrsfs}
\usepackage{amssymb}
\usepackage{bbold}
\usepackage{physics}
\usepackage{cancel}
\usepackage{slashed}
\usepackage{scalerel}
\usepackage{comment}
\usepackage{caption}
\usepackage{subcaption}
\usepackage[most]{tcolorbox}
\usepackage{xargs}
\usepackage{color}
\usepackage{float}
\usepackage{enumitem}
\usepackage[normalem]{ulem}
\usepackage{caption}
\usepackage{subcaption}
\usepackage{tikz,pgfplots}
\newcommand{\mc}[1]{\mathcal{#1}}

\newcommand{\zrh}{z_\text{rh}}

\title{\boldmath ALP Production from Abelian Gauge Bosons: Beyond Hard Thermal Loops 
}
\preprint{MITP-25-006}
\author[d,e,a]{Mathias Becker,}
\author[a]{Julia Harz,}
\author[b,c]{Enrico Morgante,}
\author[a]{Cristina Puchades-Ibáñez,}
\author[a]{Pedro Schwaller}

\affiliation[a]{PRISMA$^+$ Cluster of Excellence \& Mainz Institute for Theoretical Physics,\\
 Johannes Gutenberg-Universit\"{a}t Mainz, 55099 Mainz, Germany}
 \affiliation[b]{Dipartimento di Fisica, Università di Trieste,\\ Strada Costiera 11, I-34151 Trieste, Italy}
\affiliation[c]{INFN, Sezione di Trieste,\\ Via Valerio 2, 34127 Trieste, Italy}
\affiliation[d]{Dipartimento di Fisica e Astronomia, Universit\`a degli Studi di Padova, Via Marzolo 8, 35131 Padova, Italy}
\affiliation[e]{
INFN, Sezione di Padova, Via Marzolo 8, 35131 Padova, Italy}

\emailAdd{mathias.becker@unipd.it}
\emailAdd{julia.harz@uni-mainz.de}
\emailAdd{enrico.morgante@units.it}
\emailAdd{crpuchad@uni-mainz.de}
\emailAdd{pedro.schwaller@uni-mainz.de}

\abstract{
Previous computations of feebly interacting particle production have encountered issues with unphysical (negative) interaction rates at soft momenta. We address this problem by studying the production of Axion-Like Particles (ALPs) coupled to 
$U(1)$-gauge fields, employing the full form of 1PI-resummed gauge boson propagators. This approach avoids the need for matching or subtraction procedures, ensuring physically consistent results. We find that the ALP production rate remains positive across all momentum scales and identify the dominant production mechanisms. At soft ALP momenta 
($p \lesssim g^2 T$), interactions involving two spacelike gauge bosons dominate the production rate, surpassing other channels by an order of magnitude. 
In particular, using the full gauge boson propagator suggests that at even softer momenta ($p \lesssim g^4 T$), production involving two timelike gauge bosons becomes significant, potentially exceeding other contributions by another order of magnitude. Using these insights, we update the thermal ALP abundance and refine the estimate of the average ALP momentum, providing important input for structure formation constraints on ALP dark matter in the keV mass range.
}

\begin{document}
\setcounter{tocdepth}{1}
\maketitle

\section{Introduction}\label{sec:Intro}
Among the key quantities of theoretical astroparticle physics are particle interaction rates in the early universe. 
These rates are fundamental in determining observables such as the relic density of dark matter (DM), the baryon asymmetry of the universe, the number of relativistic degrees of freedom, and the expected thermal background of gravitational waves. 
The precise calculation of these rates, however, can be technically challenging. 
For instance, for effectively massless particles, infrared (IR) divergences arise. 
Moreover, at finite temperatures, a naive perturbative approach becomes unreliable when soft momenta\footnote{When we speak of soft momenta in the context of a gauge theory with a gauge coupling $g$, we refer to momenta of $p \sim g T$.} are involved. 
These problems can be addressed by the so-called Hard Thermal Loop (HTL) resummation of soft particle propagators (and vertices) appearing in the production process, as was realized over three decades ago~\cite{Braaten:1989mz,Braaten:1991gm,Braaten:1991we,PhysRevLett.66.2183}. 
However, the full interaction rate involves both hard and soft particle propagators, meaning the HTL resummation is not reliable everywhere. 
A simple approach to resolve this issue is to introduce a matching scale $k_\star$ with $g T \ll k_\star \ll T$, below which HTL-resummed propagators are used, while above, bare propagators are sufficient~\cite{PhysRevLett.66.2183, Bolz:2000fu, Graf:2010tv}. 
While the approach seems reliable for hard momenta, it can result in negative interaction rates at momenta below the soft scale ($p < g T$), which is clearly unphysical~\cite{Braaten:1991we,Bolz:2000fu,Pradler:2006qh,Rychkov:2007uq,Graf:2010tv,Salvio:2013iaa,Brandenburg:2004du,Baumholzer:2020hvx,Ghiglieri:2020mhm}.

Our motivation to address this problem is twofold: 
(i)~At the fundamental level, it is highly desirable to obtain a formalism that yields physical production rates for arbitrary particle momenta. After all, the momentum distribution is observable if the produced particles do not thermalize afterward, such as in freeze-in DM scenarios. 
(ii)~On the practical side, this distribution is crucial for obtaining structure formation limits on warm DM candidates such as sterile neutrinos and keV-mass ALPs~%
\cite{Shi:1998km,
Narayanan:2000tp,
Viel:2005qj,
Cadamuro:2010cz,
Arias:2012az,
Jaeckel:2014qea,
Yeche:2017upn,
Murgia:2017lwo,
Heeck:2017xbu,
Fonseca:2018kqf,
Garzilli:2019qki,
Im:2019iwd,
Fonseca:2020pjs,
Ballesteros:2020adh,
Dvorkin:2020xga,
DEramo:2020gpr}.
For example, to evaluate constraints from the Lyman-$\alpha$ forest on keV ALPs, the average DM momentum is an important input value~\cite{Baumholzer:2020hvx}. 
It can differ significantly from that of a thermal distribution and is also sensitive to contributions from the soft momentum region. 
Furthermore, the precise form of the phase-space distribution in momentum space might be relevant for predictions of the amount of dark radiation; see, for instance, \cite{DEramo:2023nzt,Bouzoud:2024bom,DEramo:2024jhn}.

Recently, Ref.~\cite{Bouzoud:2024bom} addressed the problem of negative rates in the context of thermal ALP production beyond the simple matching scheme described above. 
They proposed an improved subtraction scheme, based on methods from~\cite{Besak:2012qm,Ghiglieri:2016xye}, which mitigates and delays the onset of negative rates. Additionally, they introduced a tuned mass scheme, inspired by~\cite{Arnold:2002zm,Arnold:2003zc}, which assigns a mass proportional to the Debye mass to problematic $t$-channel propagators. A suitable multiplicative prefactor is chosen to match the HTL-resummed result in the small coupling and large ALP momentum limit, ensuring positive interaction rates for all momenta.

In this article, we follow a different approach based on the full form of the 1PI-resummed propagators instead of relying on the HTL-approximated versions. 
This method has been applied to various particle production scenarios in the context of DM~\cite{Rychkov:2007uq,Salvio:2013iaa,DEramo:2021psx,DEramo:2021lgb,becker2023dark,Eberl:2020fml,Eberl:2024pxr} and leptogenesis~\cite{Garbrecht:2013gd,Garbrecht:2013bia}. 
It allows us to use the resummed propagators at every momentum scale, avoiding any matching or subtraction procedure. 
This convenience comes at a cost: in general, the full form of the 1PI-resummed propagator, unlike the HTL-resummed version, is gauge-dependent. 
However, this problem is absent if only Abelian gauge theories are considered. 
For this reason, our analysis deals with particle production from interactions with Abelian gauge bosons. 
Specifically, we choose the frequently discussed scenario of an Axion-Like Particle (ALP) interacting with the hypercharge gauge boson to illustrate our findings.

The other key difference in our method of treating the negative rate compared to previously discussed methods is that we consider resummations for both gauge boson propagators appearing in the ALP self-energy.\footnote{The ALP production is proportional to the imaginary part of the ALP self-energy, which, at leading order in the expansion of the DM self-energy, is given by a gauge boson loop.} 
This is necessary, as we determine the production rate of soft ALPs. 
In contrast to the case of hard ALPs, both gauge bosons can be soft, requiring resummation. 
Our approach can also be motivated from first principles in the 2PI-effective action approach to non-equilibrium phenomena~\cite{Cornwall:1974vz,Chou:1984es,Berges:2004pu,Berges:2004yj}. 
There, the leading term in the loop expansion of the ALP self-energy is exactly given by the gauge boson loop with both propagators being 1PI-resummed, and this is what we aim to calculate here.

In this article, we present two results for the ALP production rate. 
First, we discuss the rate emerging from using HTL-resummed propagators for all momenta, which, of course, is inaccurate for hard gauge boson momenta. 
Afterward, we perform the calculation using the full form of the 1PI-resummed propagator, finding that the dominant double spacelike gauge boson contribution from the HTL-approximated result persists. 
However, contributions from two timelike gauge bosons, strongly suppressed in the HTL approximation, can overcome the double spacelike contribution by another order of magnitude.

The paper is structured as follows: 
In Sec.~\ref{sec:Model}, we introduce the ALP model considered in this work. 
In Sec.~\ref{sec:AxionProduction}, we present the form of the ALP self-energy and discuss the gauge boson self-energies in their different forms. 
In Sec.~\ref{sec:Results}, we present results from our numerical solutions to the ALP production rate and illustrate the phenomenological impact of our findings on structure formation constraints. 
Finally, in Sec.~\ref{sec:Conclusions}, we conclude.

\section{Model Specification}\label{sec:Model}
We consider a simple Dark Matter model, in which an ALP $a$ is coupled to an abelian gauge boson via the following Lagrangian
\begin{equation}\label{eq:ALPlang}
\mathcal{L}= \frac{1}{2}\partial_{\mu}a\partial^{\mu}a
+\frac{1}{2}m_a^2a^2 -\frac{1}{4}B_{\mu\nu}B^{\mu\nu}
-\frac{c_{i} \alpha_i}{8 \pi f_a}aB_{\mu\nu}\tilde{B}^{\mu\nu}  \, .
\end{equation}
Here, $\alpha_i= g_i^2/(4 \pi)$ is the gauge coupling, $B_{\mu\nu}$ indicates the field strength of the abelian gauge field and $\tilde B = \frac{1}{2} \epsilon B$ is its dual. In the rest of the paper, we will refer to the latter as ``photon''. When referring to the thermal vector mass, we will write $m_V$. In fact, we will mostly have in mind the gauge boson $B_\mu$ of the unbroken hypercharge $\mathrm{U}(1)_Y$ symmetry and thus will set $\alpha_i = \alpha_1$ and $c_i=c_1$ in the following. The reason for this is dictated by phenomenology: it is a known fact that, in order for our ALP to match the observed DM relic abundance, it will mostly be produced at very high temperatures, much above the EW scale~\cite{Baumholzer:2020hvx}.%
\footnote{The production in this model is UV dominated, due to the dimension-5 operator mediating interactions with the SM plasma, and thus the abundance is proportional to the reheating temperature $T_\text{rh}$, which we assume to be larger than the electroweak scale. Note that, in order to comply with indirect detection bounds, the ALP needs to have a coupling of order $c_{1} \alpha_1/ (8 \pi f_a) \sim 10^{-18}\,\mathrm{GeV}^{-1}$ and the reheating temperature should not be far from $M_\text{Pl}$~\cite{Baumholzer:2020hvx}. This constrain does not apply when the ALP couples to an abelian gauge boson of a dark sector instead of SM gauge fields or if not all of observed DM relic density is produced via ALP freeze-in. All of our findings apply regardless of the specific scenario.}
Nevertheless, we stress that the discussion of this paper is independent of this choice, and could be applied as well to the SM photon or any other dark $\mathrm{U}(1)$ field.
Other couplings to the SM fields are possible, but we choose to ignore them.
ALP couplings to non-abelian gauge bosons are for instance discussed in~~\cite{Bouzoud:2024bom,Graf:2010tv,Salvio:2013iaa,DEramo:2021psx} and ALP couplings to fermions can be found for example in \cite{Baumholzer:2020hvx,DEramo:2021usm}. 

As mentioned in the introduction, the ALP photon coupling may become problematic when computing production rates: when hard and soft regions are treated separately and then matched together as in Refs.~\cite{Braaten:1991we,Bolz:2000fu}, negative interaction rates appear at low momenta, a situation which is clearly unphysical~\cite{Baumholzer:2020hvx,Pradler:2006qh,Graf:2010tv}. This is the case in which the techniques of our paper become relevant, and we thus limit ourselves to this scenario.
Note, however, that this unphysical behavior also arises in the context of graviton \cite{Ghiglieri:2020mhm} or sterile neutrino \cite{Ghiglieri:2016xye} production from the thermal bath.

The phenomenologically relevant quantity when describing the time-evolution of particle species in the early universe is the distribution function $f(p,t)$ that depends on the absolute value of the ALP three momentum $p = |\vec{p}|$ and time $t$.
The time evolution of the distribution function is goverened by a Boltzmann-type equation
\begin{equation}
    \left[\frac{\partial}{\partial t}-H p\frac{\partial}{\partial p} \right ] f(p,t)=\mathcal{C}(p) \, .
    \label{eq: Boltz}
\end{equation}

Here $H$ is the Hubble constant and $\mathcal{C}(p)$ it is the collision term. 
It can be related to the the self energy of the particle species under consideration via \cite{Bellac:2011kqa,becker2023dark,Bodeker:2015exa} %

\begin{equation}
    \mathcal{C}(p)=\frac{\Pi^<(P)}{2 p_0}=\frac{\epsilon(p_0)}{p_0(1-e^{p_0/T})}\Im\Pi^R(P),\label{eq:collisionTerm}
\end{equation}
where $P^\mu = (p^0, \Vec{p})$ is the ALP four momentum which satisfies the on-shell condition ${p^0=\sqrt{p^2 + m_a^2}}$. 
Furthermore, $\epsilon (p^0)$ is the sign function and we use the $(+,-,-,-)$ signature for the metric tensor throughout this work. 
The non-time ordered (Wightman) self energy of the ALP $\Pi^<$ can be derived from a functional derivative of the 2PI-effective action with respect to the appropriate two-point function (see for instance \cite{Prokopec:2003pj} or \cite{Berges:2004yj} for more details), and is related to the imaginary part of the retarded self energy $\Pi^R$ via Eq.~\eqref{eq:collisionTerm}.

\begin{figure}[t!]\centering
   \includegraphics[width=0.3\textwidth]{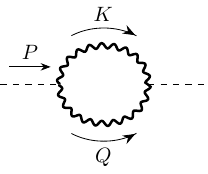} \\
   \vspace{3mm}
   \includegraphics[width=\textwidth]{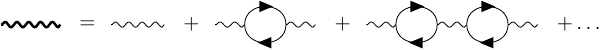}
    \caption{Feynman diagrams for the ALP self-energy $\Pi^<(P,m_a)$ and for the resummed photon propagator.}
    \label{axion-self}
\end{figure}

\begin{figure}[h]\centering
   \includegraphics[width=.8\textwidth]{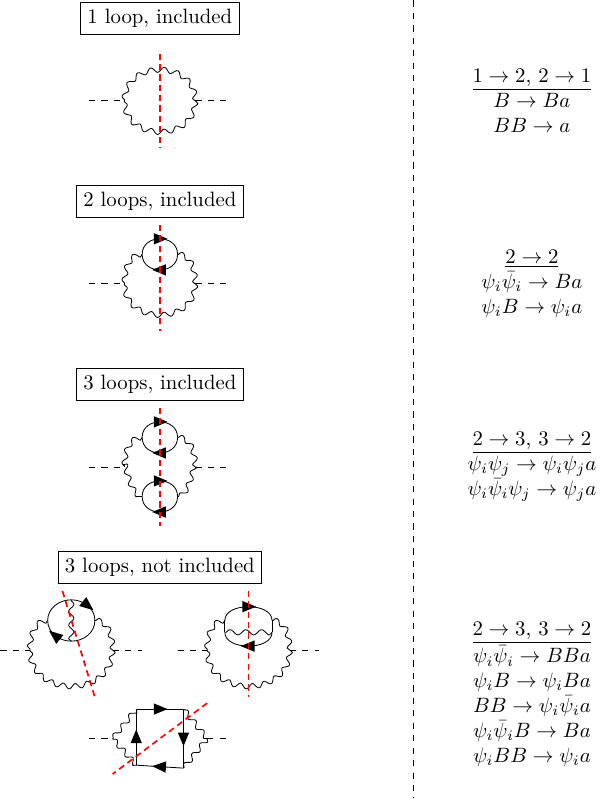}
    \caption{The first three rows illustrate the perturbative expansion of the ALP self-energy within our approximation scheme, expressed in terms of free-theory propagators up to three-loop order. These rows also depict the corresponding decay and scattering processes associated with the cuts of the ALP self-energy, all of which are thus included in our resummation scheme. In contrast, the final row shows the class of three-loop diagrams and their associated scattering processes that are not included in our calculation. Section~\ref{sec:AxionProduction} discusses the conditions under which each class of processes dominates the interaction rate. The $\psi_i$ denote all Standard Model matter fields charged under $U(1)_Y$: $\psi_i = \left \lbrace Q_L, u_R, d_R, e_L, e_R, H \right \rbrace$, where $Q_L$ and $e_L$ represent the three generations of quark and lepton doublets, $u_R$, $d_R$, and $e_R$ are the right-handed up-type quarks, down-type quarks, and leptons respectively, and $H$ is the Higgs doublet.}
    \label{fig:Cuts_and_Diagrams_included}
\end{figure}

Thus far, the collision term in Eq.~\eqref{eq:collisionTerm} is formally exact to all orders in the gauge coupling and to order $\mathcal{O} (f_a^{-2})$ in the feeble ALP coupling.
This assumes that the ALP self-energy is evaluated to all orders in its perturbative expansion, using the exact propagators of the theory. 
However, this is practically not achievable such that we employ two approximations when deriving the ALP self energy:
\begin{enumerate}
\item We truncate the perturbative expansion of the ALP self energy at leading order such that it is given by the photon-loop shown in Fig.~\ref{axion-self}.
Note that the only contribution arising at two-loop level, obtained by inserting an ALP in the photon loop, is suppressed by two powers of the feeble ALP coupling $\sim c_1 \alpha_1 T/(8 \pi f_a)$ as long as $T_\text{rh} \ll f_a$, such that the first relevant corrections to the one-loop self-energy only arise at three-loop level.
\item We approximate the photon propagators using one-loop 1PI-resummation. The one-loop photon self energy is computed using the propagators of the free theory, as indicated in the second line of Fig.~\ref{axion-self}.
\end{enumerate}
In Fig.~\ref{fig:Cuts_and_Diagrams_included}, we illustrate, up to the three-loop level using bare propagators, the classes of scattering diagrams corresponding to cuts of the ALP self-energy that are included in our calculation. 
Note that Fig.~\ref{fig:Cuts_and_Diagrams_included} does not depict all possible cuts; in particular, it omits cuts associated with virtual corrections to $1 \leftrightarrow 2$ and $2 \leftrightarrow 2$ processes.

This approximation scheme captures part of the leading-order contributions to the ALP production rate in the soft-momentum regime $g_1^4 T < p < g_1 T $, which, as discussed in Sec.~\ref{sec:AxionProduction} arises from $2 \leftrightarrow 3$ scatterings. 
For harder momenta ($p > g_1 T$), all leading-order contributions are included.
We will discuss the contributing processes in more detail in Sec.~\ref{sec:AxionProduction}.

Before we proceed to calculate $\Pi^<$ in the next section, we express Eq.~\eqref{eq: Boltz} in terms of comoving momenta $p\rightarrow p/T$ and the time-variable $z = m_a/T$. 
Under the assumption that the entropy degrees of freedom are constant, we can use $d T / d t = -H T $, and together with the definition of $\mathcal{C}(p)$ we arrive to the following relation

\begin{equation}
    \frac{\partial f(p/T,z)}{\partial z} =\frac{1}{2 p_0 z H}\Pi^<(p/T,z) \, .
    \label{eq:BoltzTFT}
\end{equation}
The preceding expression illustrates that the time evolution of each comoving momentum mode can be treated independently such that the ALP distribution function $f(p/T,z)$ can simply be found by integrating over $\Pi^{<} (p/T,z)$ according to Eq.~\eqref{eq:BoltzTFT} for each comoving momentum.
\clearpage

\section{Thermal ALP Production at Finite Temperature}\label{sec:AxionProduction}
In this section, we compute the ALP collision term according to Eq.~\eqref{eq:collisionTerm}.
The ALP Wightman self-energy $\Pi^<(P)$ is given by a gauge boson loop as illustrated in Fig.~\ref{axion-self}, where both photon propagators are 1PI-resummed.
Then, following \cite{Salvio:2013iaa}, the ALP self-energy results in
\begin{equation}\label{eq:axion-SE}
    \Pi^<(P)=\frac{c_{1}^2 \alpha^2_1}{8 f^2_a \pi^2}\int\frac{d^4K}{\left( 2\pi \right)^4}\epsilon^{\mu\nu\alpha\beta}\epsilon^{\mu'\nu'\alpha'\beta'} K_\alpha Q_\beta K_{\alpha'}Q_{\beta'} D^<_{\mu\mu'}(K) D^<_{\nu\nu'}(Q) \, .
\end{equation}

Again, $\alpha_1= g_1^2/(4 \pi)$ is the hypercharge gauge coupling. 
Moreover, $D^<_{\mu\mu'}(K)$ denotes the 1PI-resummed Wightman propagator of a gauge boson and $Q = P - K$. 
The retarded one-loop-resummed gauge boson propagator, whose imaginary part is related to the Wightman propagator via $\text{Im} (D_{\mu \nu}^{R} (K)) = - \frac{1}{2 f_B(k^0)} D_{\mu \nu}^< $, can be written as~\cite{LeBellac}:
\begin{equation}\label{eq:res-prop}
    D_{\mu \nu}^R (K) = i \left[ \frac{\mathcal{P}_{\mu \nu}^t}{K^2-\pi_0(K)-\pi_t(K)} + \frac{\mathcal{P}_{\mu \nu}^l }{K^2-\pi_0(K)-\pi_l(K)}+ \frac{\mathcal{P}^K_{\mu \nu} }{K^2}\right] \, . 
\end{equation}
Here, the polarizations are conveniently decomposed into transverse (orthogonal to $K$), longitudinal (orthogonal to $K$ and parallel to $\vec{k}$), and parallel to $K$. 
The quantity \( \pi_0 \) denotes the retarded vacuum self-energy of the gauge boson, while \( \pi_{t/l} \) represent the transverse and longitudinal components of the finite-temperature part of the retarded self-energy. 
The gauge boson Wightman propagator is then expressed through the spectral densities \( \rho_{t/l} \)~\cite{Salvio:2013iaa}
\begin{equation}\label{eq:non-time-propagator.}
    D_{\mu \nu}^< (K) = f_B (k_0) \left[ \mathcal{P}_{\mu \nu}^t\rho_t (K) + \mathcal{P}_{\mu \nu}^l \frac{k^2}{K^2} \rho_l (K) + \xi \frac{K_\mu K_\nu}{K^2} \right] \, ,
\end{equation}
where $\xi$ parametrizes the gauge dependence. 
Here, $f_B(k_0)$ is the Bose-Einstein distribution:
\begin{equation}\label{eq:BE}
    f_B(k_0)=\frac{1}{e^{k_0/T}-1}.
\end{equation} 
The spectral densities have the following form:
\begin{align}
    \label{eq:spectraldensT}\rho_t&=-2\frac{\Im\pi_0 +\Im\pi_t}{(K^2-\Re\pi_0-\Re\pi_t)^2+(\Im\pi_0+\Im\pi_t)^2} \, , \\
     \label{eq:spectraldensL}\rho_l&=-2\frac{K^2}{k^2}\frac{\Im\pi_0 +\Im\pi_l}{(K^2-\Re\pi_0-\Re\pi_l)^2+(\Im\pi_0+\Im\pi_l)^2}  \, .
\end{align}
Here we have already decomposed the thermal photon self-energies into their real and imaginary part.
By combining Eq.~\eqref{eq:non-time-propagator.} and Eq.~\eqref{eq:axion-SE}, and introducing the identity \( 1 = \int d^4 Q \, \delta(K+Q-P) \) to simplify the angular integrations, the subsequent index contractions yield the following expression~\cite{Salvio:2013iaa}

\begin{align}\label{eq:TL-axionSE}
    \Pi^< (P) = \frac{C_{\Pi}}{f_a^2 p} \int_{-\infty}^\infty dk^0 \int_0^\infty dk \int_{|p-k|}^{|p+k|} dq \, k q f_B (k^0) f_B(p^0 - k^0) \left[ \mc{I}_{lt} + \mc{I}_{tt} \right] \, , 
\end{align}
where we introduced the constant 
\begin{equation}\label{eq:CPi}
    C_{\Pi}=\frac{c_{1}^2\,\alpha_1^2}{8(2\pi)^5}
\end{equation}
and
\begin{align}\label{eq:LT-TTcontrib}
    \mc{I}_{lt} &= \left( \rho_t(K) \rho_l(Q) + \rho_l(
    K) \rho_t(Q) \right) \left[ (k+q)^2 - p^2 \right] \left[ p^2 - (k-q)^2 \right] \, , \\
    \mc{I}_{tt} &= \rho_t(K) \rho_t(Q)\left[ \left( \left(\frac{k_0}{k}\right)^2  + \left(\frac{q_0}{q}\right)^2 \right) ( (k^2- p^2 + q^2)^2 + 4k^2q^2 ) + 8 k^0 q^0 (k^2 +q^2 -p^2) \right] \, .
\end{align}
The terms $\mc{I}_{lt/tt}$ refer to the longitudinal-transverse and double-transverse polarization modes of the two-photon contributions.

Until now, the expressions derived are independent of the specific form of the photon self-energy.
Importantly, we find that the expression in Eq.~\eqref{eq:TL-axionSE} is gauge independent, since the gauge-dependent polarization parallel to the photon four-momentum cancels and the longitudinal and transverse components of the self-energy tensors $\pi_{l/t}$ are as well gauge independent. 

In this paper, we perform the calculation in two ways: first using HTL-approximated propagators, and then using the full form of the 1PI-resummed propagators.
While the resulting HTL propagators offer a simpler analytic form and are easier to compute, they are not reliable for hard momenta, requiring the use of matching or subtraction schemes to ensure accuracy. 
Previous works have used methods such as mixed propagators, HTL resummed for soft momenta, and free propagators for hard momenta to address this limitation~\cite{Braaten:1991we,Bolz:2000fu,Pradler:2006qh,Rychkov:2007uq,Graf:2010tv,Salvio:2013iaa,Brandenburg:2004du,Baumholzer:2020hvx,Ghiglieri:2020mhm}. 
However, this procedure results in the negative ALP production rate.
On the other hand, calculating the one-loop photon self-energy without approximation—what we refer to as the `1PI' case in the following—is valid and accurate across all momentum scales.
Thus, we expect the full form of the 1PI-resummed propagator to result in a more reliable prediction for particle production rates from abelian gauge bosons.
For comparison we also give the results obtained using HTL-resummed propagators even outside their range of validity. 

To illustrate the differences between the spectral functions in Eqs.~\eqref{eq:spectraldensT} and~\eqref{eq:spectraldensL} under the HTL approximation, we examine the photon self-energy, which has been extensively studied in the literature (e.g.,~\cite{PhysRevD.26.1394,Rychkov:2007uq,Drewes:2024wbw}):

\begin{subequations}
\label{eq:photonSE}
\begin{align}
\pi_l (K)&=-\frac{K^2}{k^2}g_1^2 \left( N_S H_S(K)+N_F H_F(K) \right)\simeq - 2 m^2_V\frac{K^2}{k^2}\left( 1-\frac{k_0}{2k}\log{\frac{k_0+k}{k_0-k}}\right),\label{eq:PhtonSEL}\\
\pi_t (K) &=-\frac{\pi_l (K)}{2}+\frac{g_1^2}{2} \left( N_S G_S(K)+N_F G_F(K) \right) \simeq m^2_V\frac{k_0^2}{k^2}\left( 1-\frac{k_0}{2k}\log{\frac{k_0+k}{k_0-k}}\right).\label{eq:PhtonSET}
\end{align}
\end{subequations}
The first equality corresponds to the full 1PI-resummed self-energies, without any approximation. 
The symbol $\simeq$ indicates the use of the HTL approximation. 
The coefficients $N_S = \sum_{i \in s} Y_i^2 = 1/2$ and $N_F = \sum_{i \in f} Y_i^2 = 10$ encode the number of scalar (S) and fermionic (F) degrees of freedom in the SM weighted by their hypercharge $Y_i$ and enter in the thermal vector mass $m_V^2=\frac{1}{6}g_1^2T^2(N+N_S+N_F/2)$.
The explicit forms of the integrals $H_{S/F}(K)$ and $G_{S/F}(K)$ are provided in Eqs.~\eqref{ImHS}, \eqref{ImHF}, \eqref{ImGS}, and \eqref{ImGF} for their imaginary parts, and in Eqs.~\eqref{ReHS}, \eqref{ReHF}, \eqref{ReGS}, and \eqref{ReGF} for their real parts. Furthermore, the retarded vacuum self-energy $\pi_0$ is detailed in Eq.~\eqref{eq:ImVaacSE}.
For the HTL computation we can neglect this contribution. 
However, it is important that we include the vacuum contribution when using the full form of the self-energies.%
\footnote{Practically, we only include the imaginary part of the vacuum photon self-energy $\pi_0$ in our numerical calculations, which is necessary to ensure that the photon spectral function obeys $\text{sign} \left( \rho_{l/t} \left( K \right) \right) = \text{sign} \left( k^0 \right)$. We neglect the real part of $\pi_0$ that has to be renormalized as usual at $T=0$. We have verified that it has a negligible impact on the results when the renormalization scale is identified with the temperature $\mu=T$.}

In the following, we exploit the simpler form of the HTL self-energy to discuss some qualitative features of ALP production from abelian gauge bosons and provide an overview of the conceptual differences between the two methods. 
Note that the imaginary part of the photon self-energy in Eq.~\eqref{eq:photonSE} is obtained from the analytic continuation $k_0 \rightarrow k_0 + i\epsilon$. For the transverse contribution, we find
\begin{align}\label{eq:HTLSETransv}
    \Im\pi_t &\simeq \frac{m_V^2}{2}\frac{K^2}{k^2}\frac{k_0}{k}\pi\Theta(-K^2)  & 
   \Re\pi_t &\simeq m^2_V\frac{k_0^2}{k^2}\left( 1-\frac{k_0}{2k}\log{\frac{k_0+k}{k_0-k}}\right) ,
\end{align}
while the longitudinal component results in
\begin{align}\label{eq:HTLSELong}
    \Im\pi_l &\simeq -\frac{m_V^2}{2}\frac{K^2}{k^2}\frac{k_0}{k}\pi\Theta(-K^2)  & 
   \Re\pi_l &\simeq -2 m^2_V\frac{K^2}{k^2}\left( 1-\frac{k_0}{2k}\log{\frac{k_0+k}{k_0-k}}\right) \, .
\end{align}
From these expressions, we observe an important fact: due to the form of the logarithmic contribution in Eq.~\eqref{eq:photonSE}, the imaginary parts of the self-energies are non-zero only in the regime $K^2 < 0$, or equivalently in the spacelike regime.

\begin{figure}[t!]
 \begin{subfigure}{0.49\textwidth}
     \includegraphics[width=\textwidth]{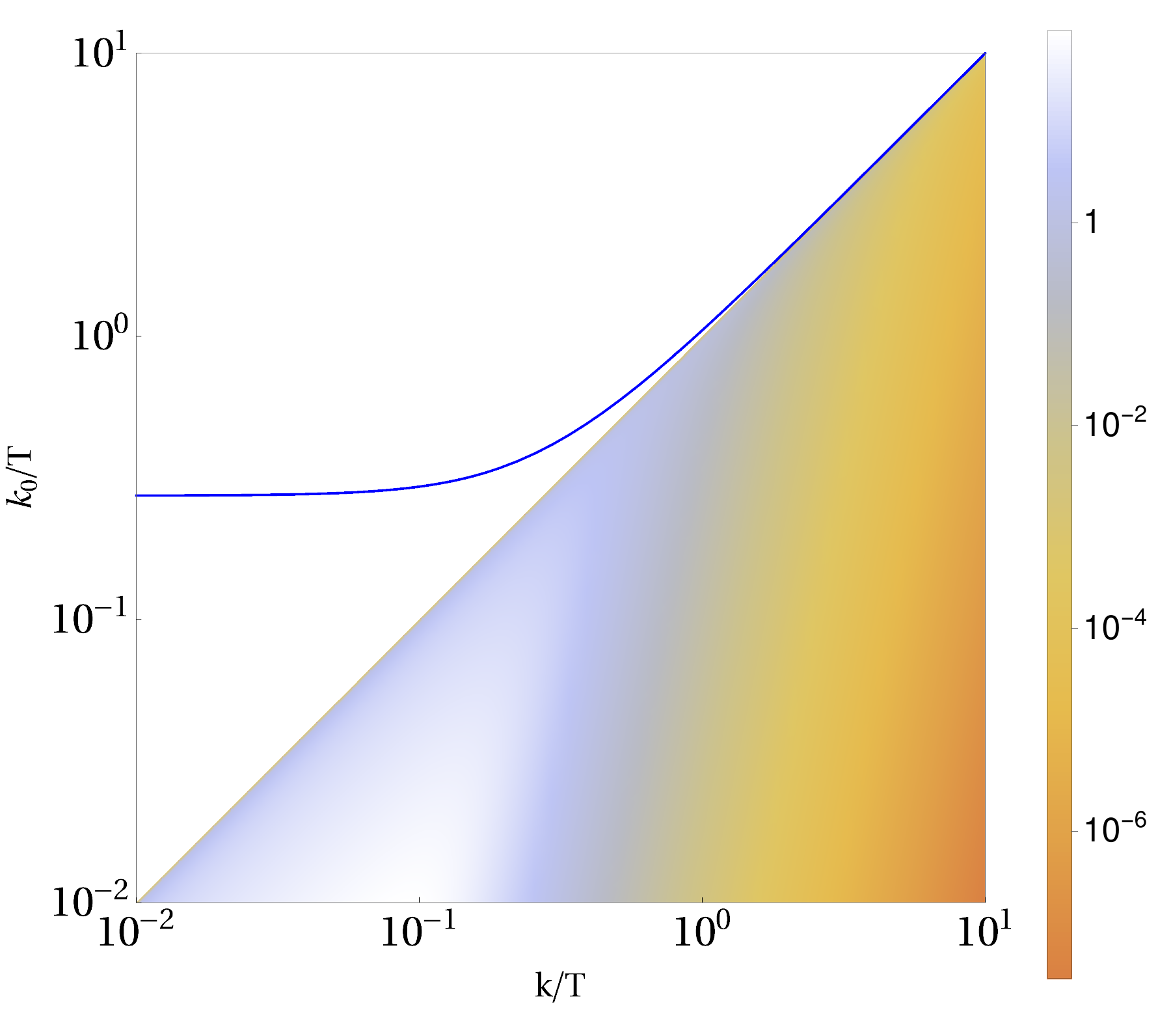}
     \caption{HTL transversal }
     \label{fig:DensHTLTr}
 \end{subfigure}
 \hfill
 \begin{subfigure}{0.49\textwidth}
     \includegraphics[width=\textwidth]{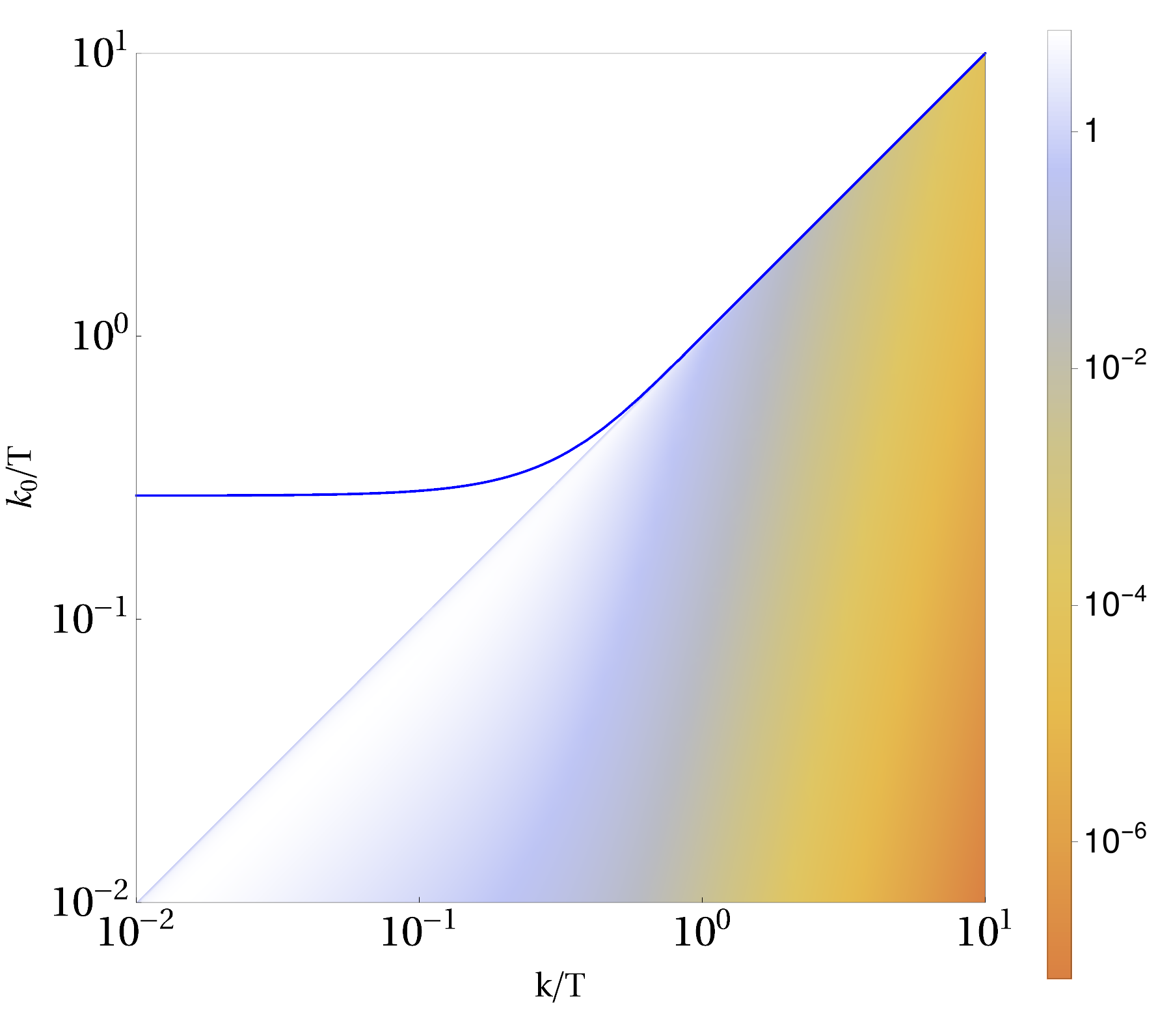}
     \caption{HTL longitudinal }
     \label{fig:DensHTLlg}
 \end{subfigure}
 
 \medskip
 \begin{subfigure}{0.49\textwidth}
     \includegraphics[width=\textwidth]{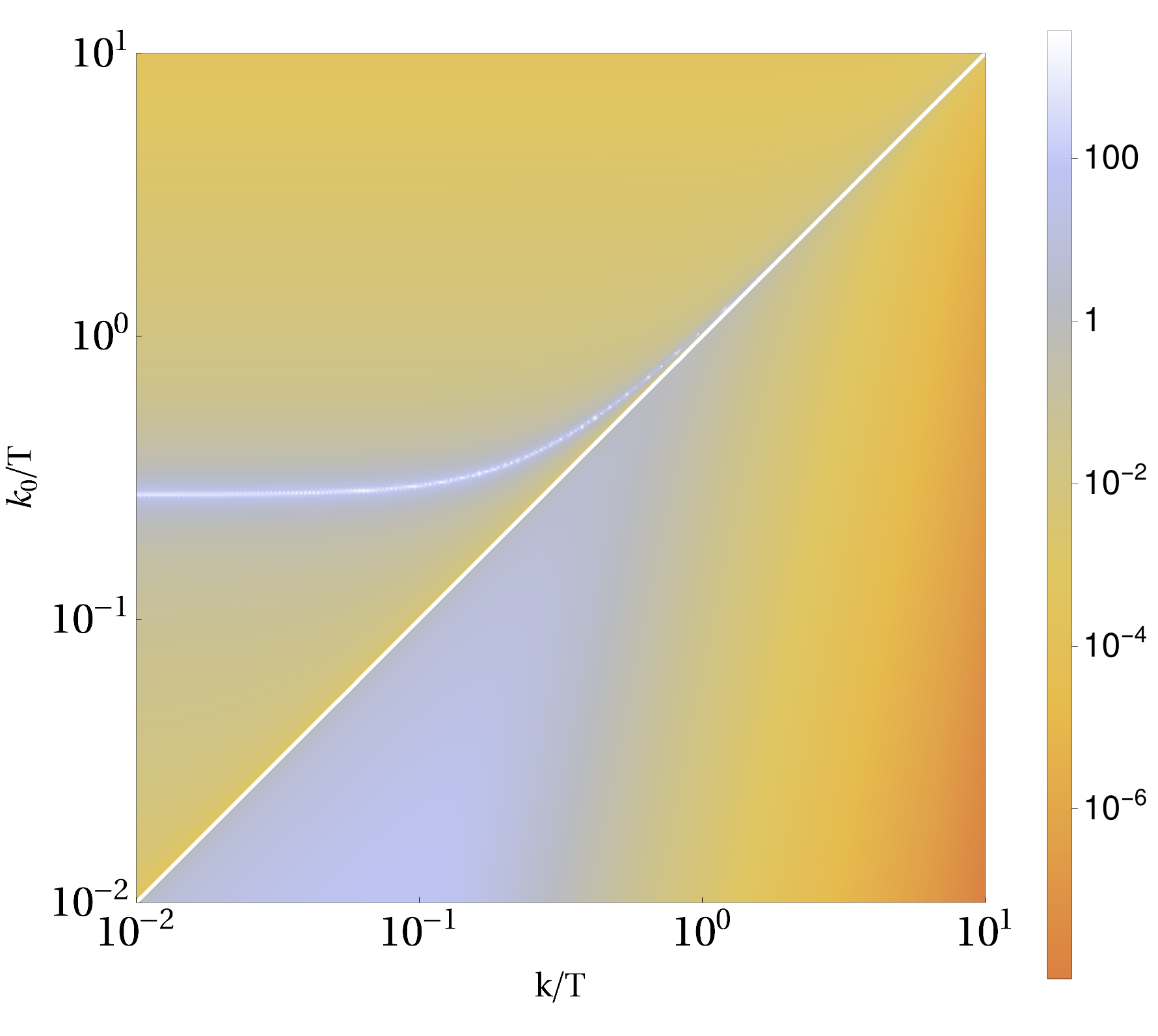}
     \caption{1PI transversal }
     \label{fig:DensFullTr}
 \end{subfigure}
 \hfill
 \begin{subfigure}{0.49\textwidth}
     \includegraphics[width=\textwidth]{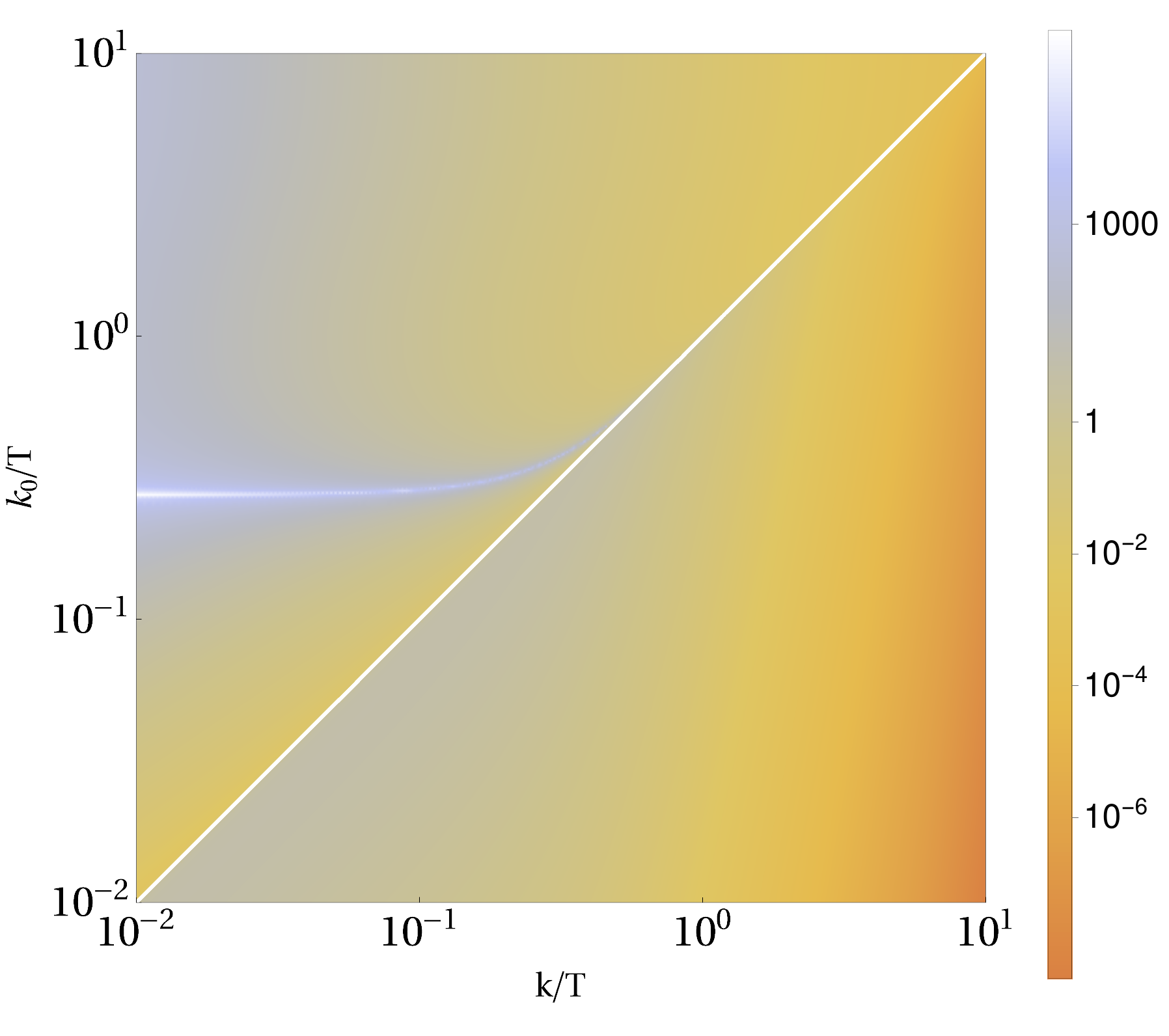}
     \caption{1PI longitudinal}
     \label{fig:DensFullLg}
 \end{subfigure}

 \caption{The left column illustrates the transversal 1PI-resummed photon spectral density $\rho_t$, Eq.~\eqref{eq:spectraldensT}, in the HTL approximation (top) and its full form (bottom). The right column shows  the longitudinal 1PI-resummed photon spectral density $\rho_l$, Eq.~\eqref{eq:spectraldensL}, in the HTL approximation (top) and its 1PI-resummed form (bottom). The HTL spectral densities are given by $\delta$-functions at timelike momentum indicated by the blue line. We set the hypercharge coupling $g_1=0.35$.}
 \label{fig:SpectDens}

\end{figure}

The fact that the HTL imaginary part vanishes for timelike momentum, i.e. $K^2>0$, translates into a well-defined particle pole in this regime, similar to the behavior in vacuum but displaced from the lightcone by the presence of a thermal mass.  
This can be observed in Fig.~\ref{fig:SpectDens}. 
The key difference between the HTL spectral densities and those from the 1PI-resummed propagator is that the clearly defined pole in the HTL approximation acquires a width in the full theory and a continuum structure emerges throughout the spectrum, which relaxes kinematic constraints that are present for the HTL spectral density if only timelike gauge bosons are involved in the production process.
For spacelike momentum the structure of the HTL approximated and the full form of the 1PI-resummed spectral densities are similar and feature a continuum, which mainly differs at large spacelike momenta $k \gg g T$ and alter the rate of hard ALPs that are not the primary interest of this study. 
For example, this difference is quantified in \cite{becker2023dark} where a feebly interacting DM candidate produced via decays or scatterings of a massive gauge-charged parent particle is analyzed.

The clear distinction within the HTL spectral density between the timelike and spacelike momenta significantly impacts the expected results and also gives us an indication of what to expect when considering the full form of the 1PI-resummed propagators. 
Therefore, we must study the following three regions separately:
\begin{itemize}
    \item Timelike-Timelike (TT) interaction: Both photon momenta are timelike, interpreted as a particle-particle interaction.
    \item Timelike-Spacelike (TS) interaction: One photon momentum is timelike and the other is spacelike, interpreted as a particle-plasma interaction. This interaction has two symmetric contributions.
    \item Spacelike-Spacelike (SS) interaction: Both photon momenta are spacelike, which can be interpreted as a plasma-plasma interaction.
\end{itemize}

In the following, we discuss these regimes with a special emphasis on the small ALP momentum limit $p \ll m_V$, where we provide an estimate of the scaling of each individual contribution with $p$.\\

\textbf{Timelike-Timelike (TT) contributions:} 
In the HTL approximation, the gauge boson propagator has a vanishing width, which relates this regime to $1 \rightarrow 2$ processes, as shown in the first line of Fig.~\ref{fig:Cuts_and_Diagrams_included}. 
The gauge bosons acquire a momentum-dependent thermal mass, differing for transverse and longitudinal components. 
In the large momentum limit, the transverse thermal mass is given by $m_T = m_V$, while the longitudinal components mass is exponentially suppressed,  $m_L \approx 0$.  
Since ALP freeze-in via gauge bosons proceeds at high temperatures, the ALP mass can be neglected.
Therefore, $1 \rightarrow 2$ proceses involving two transverse gauge bosons and an ALP are kinematically forbidden in the HTL approximation. 
Decays of a transverse gauge boson into an ALP and a longitudinal gauge boson are possible but only relevant at temperatures close to the ALP mass, which is not important for this discussion.

The situation is fundamentally different when considering the full form of the photon self energies, as they develop a finite imaginary part such that the kinematic constraints are relaxed. 
Assuming $p \ll m_V$, we can find an approximate solution to Eq.~\eqref{eq:TL-axionSE}, which for the dominant contribution from two transverse gauge bosons yields
\begin{align}
    \Pi_{TT}^< (p) = C_{TT} (m_V) \begin{cases}
          p^0  & p \gtrsim p_c \\
         p^2 & p \lesssim p_c
    \end{cases}  \, ,
\end{align}
where $p_c$ marks the momentum scale where the scaling behavior of $\Pi_{a,TT}^<$ changes and it can be estimated as the gauge boson width at zero momentum $\Gamma_{TT} = \text{Im} \Pi / k^0$ at the particle pole, i.e. $k^0 (k=0) \approx \sqrt{2/3} m_V$.
Moreover, $C_{TT} (m_V)$ is a prefactor detailed in the appendix that contains a one-dimensional integral that needs to be solved numerically. 
As long as $p_c < p \ll m_V$, the ALP self-energy from TT-type contributions is momentum independent, while all other contributions are at least power-law suppressed with $p$.
Hence, as we demonstrate in our numerical results, TT-type contributions can constitute the dominant part of the ALP interaction rate for soft ALP momenta. 
This kinematic regime can be identified with s-channel $2 \leftrightarrow 2$ or $2 \leftrightarrow 3$ scatterings, as illustrated in Fig.~\ref{fig:Cuts_and_Diagrams_included}. \\

\textbf{Timelike-Spacelike (TS) contributions:} At hard ALP momenta and large temperatures, TS-type contributions, which can be identified with t-channel $2 \leftrightarrow 2$ scatterings (second line if Fig.~\ref{fig:Cuts_and_Diagrams_included}), dominate the ALP interaction rate.  
On the contrary, for small ALP momenta $p \ll m_V$, their contribution is exponentially suppressed. 
Using HTL approximated gauge boson propagators, we can estimate the ALP self-energy as
\begin{align}
    \Pi_{TS}^< (p) 
    \approx  \frac{C_{\Pi} T}{f_a^2 } 16 \pi^2  m_V^2 p  \exp \left( - \frac{m_V^2}{4p T} \right) \, .
\end{align}
This result can safely be applied when considering the full form of the propagators, as it is dominated from contributions arising from two almost lightlike gauge bosons, where HTLs are accurate. 
Consequently, this contribution is negligible when considering the production of ALPs with momenta $p \ll m_V$. \\

\textbf{Spacelike-Spacelike (SS) contributions:} At hard ALP momenta, SS-type contributions, which can be associated with t-channel $2 \leftrightarrow 3$ scatterings (third line in Fig.~\ref{fig:Cuts_and_Diagrams_included}), are subleading to TS-type contributions due to a larger suppression with the gauge coupling. 
However, at momenta $p \ll m_V$, we find an approximate solution of the ALP self-energy using HTL resummed propagators
\begin{align}
    \Pi_{SS}^< (p) &\approx 93 \frac{C_\Pi T^4}{f_a^2} m_V^\frac{2}{3} p^\frac{4}{3} \, ,
\end{align}
where the numerical coefficient is chosen to reflect our numerical findings while the scaling with $p$ and $m_V$ can be found analytically, as we detail in Appendix \ref{app:B}. 
Our findings also apply to the case where the full form of the gauge boson propagator is considered, as this contribution is dominated by gauge bosons with momenta $k \ll m_V$ where HTLs are reliable.
As a result, SS-type contributions dominate the ALP interaction rate at soft momentum when HTL resummed propagators are considered. 
Note that a similar effect was also found in Ref.~\cite{Laine:2011xm}, see Fig.~3 therein.
However, for the full form of the 1PI-resummed propagator, SS-type contributions can be subleading compared to TT-type contributions as long as $p \gtrsim p_c$. 
Below this momentum scale they eventually constitute the dominant contribution again. 
\newline

Before proceeding to our numerical results, we would like to comment on the shortcomings and potential improvements of our work. 
Above we have pointed out that $2 \leftrightarrow 3$ scatterings can constitute the leading order contribution to the ALP interaction rate at soft momenta $p \lesssim m_V^2 / T$.
These scatterings arise first at three-loop level in a perturbative expansion of the ALP self-energy.
As we illustrated in Fig.~\ref{fig:Cuts_and_Diagrams_included}, our 1PI resummations do not include all classes of three-loop diagrams corresponding to $2 \leftrightarrow 3$ scatterings such that we only capture parts of the leading order contribution. 
Specifically, we do include $2 \leftrightarrow 3$ scatterings involving soft ALP emission from virtual photons, e.g. $f f \rightarrow f f a$, as illustrated in the second line of Fig.\ref{fig:Cuts_and_Diagrams_included}.
However, we do not include soft ALP emission from external photons, for instance the process $f \gamma \rightarrow f \gamma a$, which correspond to two-loop corrections to the photon self-energy or vertex corrections of the ALP self-energy at three-loop level as illustrated in the last line of Fig.~\ref{fig:Cuts_and_Diagrams_included}.
This is similar to the case of soft dilepton emission, where Ref.~\cite{Moore:2006qn} noted that the 1PI resummations in Ref.~\cite{Braaten:1990wp} overlooked this contribution.
However, while we acknowledge the importance of these class of diagrams, their inclusion would lead to significantly more complex numerical calculations.
Hence, we have decided to leave this ambitious extension for future work.
\section{Results}\label{sec:Results}
In this section, we present our results on the ALP self-energy $\Pi^<$ in Eq.~\eqref{eq:axion-SE}, which is directly related to the ALP momentum distribution $f(p)$ (see Eq.~\eqref{eq:BoltzTFT}).  
For our analysis, we set the hypercharge gauge coupling to $g_1=0.35$.%
\footnote{As the primary objective of this work is to investigate particle production rates at soft momentum and identify the dominant kinematic regimes, we neglect the running of the couplings involved.}

\begin{figure}[t!]
    \centering\includegraphics[width=0.67\textwidth]{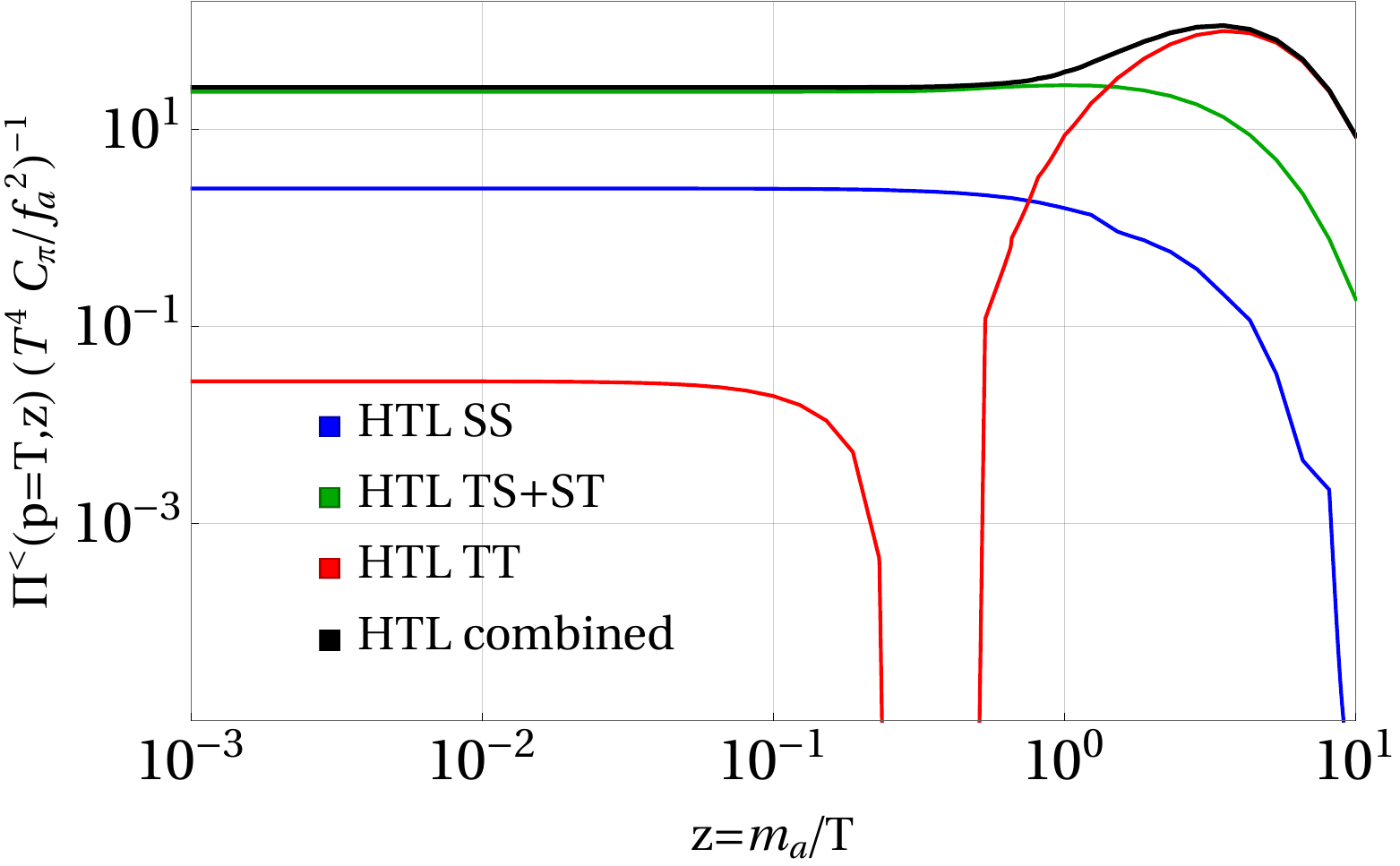}
    \caption{Wightman self-energy of the ALP $\Pi^<(p,z)$ using the HTL approximation for fixed ALP momentum $p=T$. We can observe how all the different contributions tend to follow a constant when $m_a<p$.}
    \label{fig:AxDistribHTL}
\end{figure}

We begin by discussing the ALP self-energy $\Pi^<(P)$ using gauge boson propagators in the HTL approximation. 
After performing all the integrals in Eq.~\eqref{eq:TL-axionSE} numerically, we obtain $\Pi^<$ as a function of the absolute value of the ALP three-momentum $p$, and the ALP mass normalized to the temperature $z=m_a/T$. 
When calculating the production rate for a fixed ALP momentum $p$, we observe that the rate approaches a constant value for $z \lesssim p$. 
This behavior is illustrated in Fig.~\ref{fig:AxDistribHTL} for each kinematic regime for a fixed momentum $p=T$. 
Since ALP freeze-in mediated by a coupling to gauge bosons proceeds via a higher-dimensional operator, the production is dominated by large temperatures around the reheating temperature.
Therefore, in our further analyses, we present results exclusively for $z=0$, as these results accurately describe the dominant contribution to the ALP distribution function,  provided $z_{rh} \ll p_\mathrm{min} / T$, where $p_\mathrm{min} / T = 10^{-3}$ represents the minimal momentum considered in our analysis.
\begin{figure}[t!]
    \centering\includegraphics[width=0.7
    \textwidth]{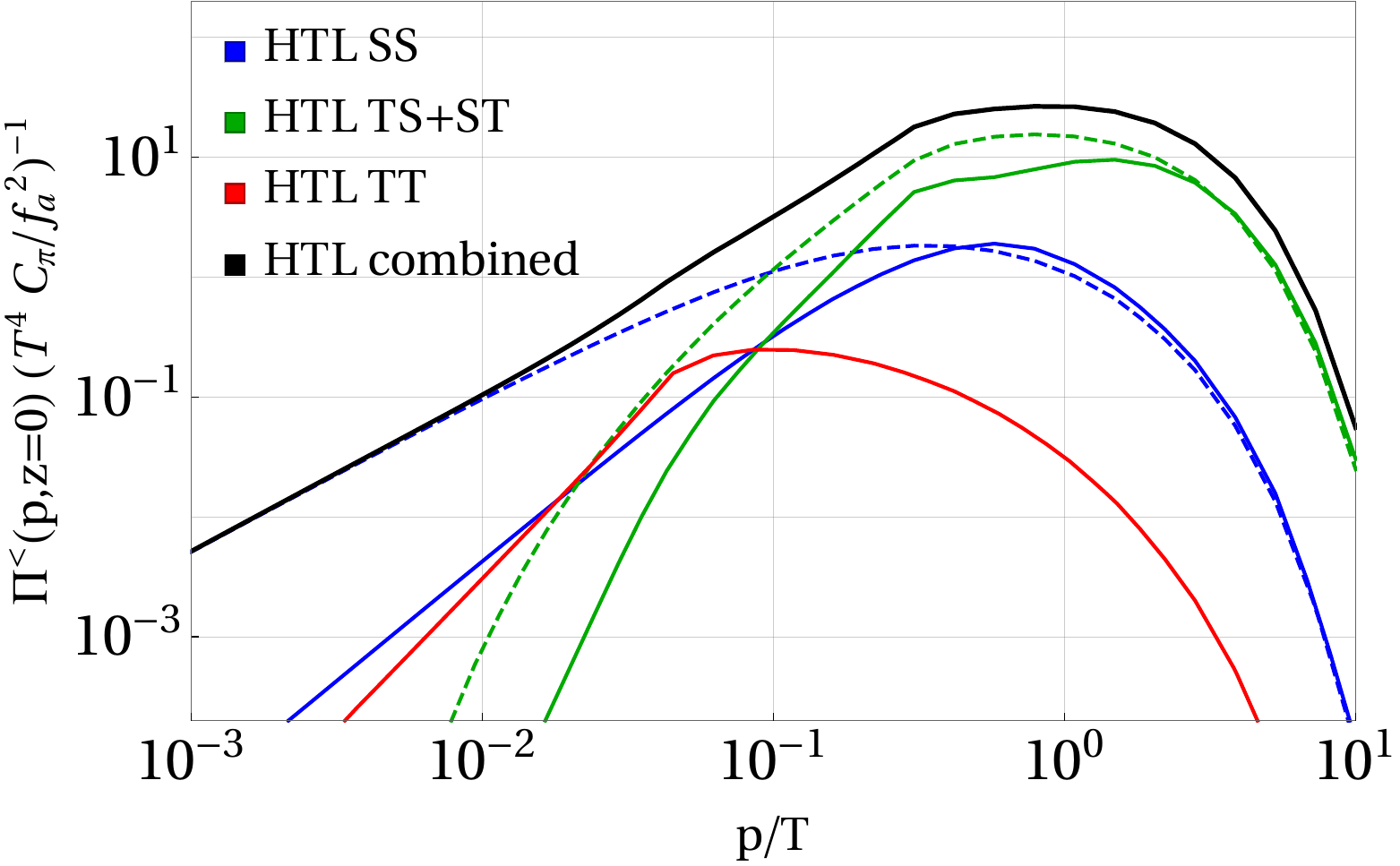}
    \caption{ALP self energy $\Pi^<$ using the HTL approximation for $z=0$. Solid lines correspond to the longitudinal-transversal ($\mathcal{I}_{lt}$) contribution and dashed lines to the double transversal contribution ($\mathcal{I}_{tt}$), see Eq.\eqref{eq:LT-TTcontrib}.}
    \label{fig:AxDistribHTLz0}
\end{figure}

\paragraph{HTL ALP self-energy.}
In Fig.~\ref{fig:AxDistribHTLz0}, we present the results for the HTL ALP self-energy at $z = 0$, which will later be compared to the results obtained using the 1PI-resummed photon propagators.  
We emphasize that positive results for the ALP production rate are obtained across all momenta, thereby avoiding the unphysical negative rates for $p < g_1 T$ previously identified when employing a matching procedure used, for instance, in~\cite{Braaten:1991we,Bolz:2000fu,Pradler:2006qh,Rychkov:2007uq,Graf:2010tv,Salvio:2013iaa,Brandenburg:2004du,Baumholzer:2020hvx,Ghiglieri:2020mhm}.  
Additionally, as expected, the contributions from two timelike (TT) photons are subleading.  
Indeed, as shown in Fig.~\ref{fig:AxDistribHTLz0}, the double transverse contribution vanishes, while the transverse-longitudinal contribution (solid red line) is always negligible.  
At hard momentum, $p \gtrsim g_1 T$, the dominant contribution arises when one photon is spacelike and the other is timelike (TS), corresponding to $2 \leftrightarrow 2$ scatterings, as represented by the green lines.  
However, at smaller momenta, TS-type contributions scale as $\Pi_a^< \sim p \exp(-m_V^2 / (pT))$ and are therefore negligible.  

Finally, at small momentum, $p \lesssim 0.1 T \sim g_1^2 T$, contributions from two spacelike photons (SS), corresponding to $2 \leftrightarrow 3$ scatterings, scale as $\Pi^< \sim p^{4/3}$ and dominate the interaction rate.  
We stress that this contribution arises exclusively when both propagators are resummed and exceeds the contributions otherwise present by more than one order of magnitude, see also~\cite{Laine:2011xm}.  
Although these results demonstrate that the interaction rate calculated using HTL-resummed propagators remains positive for all momenta, this approach becomes unreliable when hard photon momenta are involved.  
This scenario cannot be neglected, as the ALP interaction rate is determined by an integral over all photon momenta.

\begin{figure}[t!]
    \centering\includegraphics[width=0.7\textwidth]{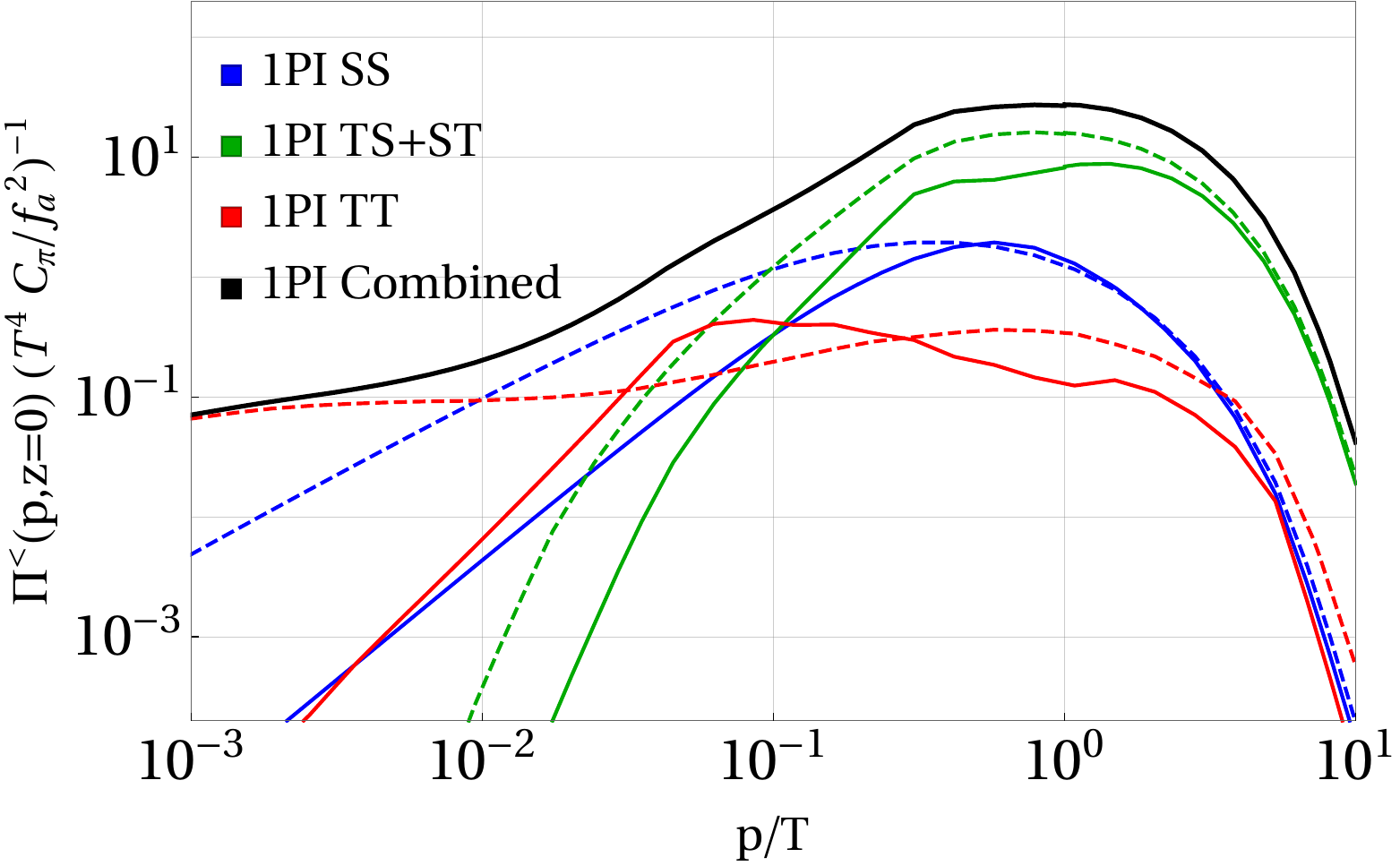}
    \caption{ALP self energy $\Pi^<$ using 1PI-resummed photon propagators for $z=0$. Solid lines correspond to the Longitudinal-Transversal ($\mathcal{I}_{lt}$) contribution and dashed lines to the double transversal contribution ($\mathcal{I}_{tt}$), see Eq.\eqref{eq:LT-TTcontrib}.}
    \label{fig:FullAxProd}
\end{figure}

\paragraph{1PI ALP self-energy.}
In Fig.~\ref{fig:FullAxProd}, we present the ALP interaction rate obtained from the 1PI-resummed photon propagators, instead of relying on methods that combine HTL-resummed and free-theory propagators.  
At hard ALP momentum, we find qualitative agreement between the ALP self-energy calculated using HTL-resummed photon propagators and the 1PI-resummed result. 
The self-energy remains dominated by TS-type contributions differing from the HTL-resummed result by $\mc{O} (10 \%)$\footnote{For large ALP momenta $p>T$, we calculated the interaction rate of TS-type contributions replacing the 1PI-resummed propagator with its HTL approximated version at timelike momentum. This was done as it simplifies the numerical integral by removing a numerical integration over a sharply peaked function. While it is possible to perform a numerical treatment that includes the 1PI-resummed propagator in this kinematic regime (as done, for example, in \cite{becker2023dark}), such a calculation is very time-consuming. Since hard ALP momenta are not the primary focus of this work, we have opted not to pursue it. We stress that for all other kinematic regimes the 1PI-resummed photon propagator has been considered in the numerical evaluation.}. 
At soft momentum, both SS-type and TS-type contributions accurately reproduce the HTL-resummed result. 
This is expected, as SS-type contributions at soft ALP momentum are dominated by soft photons, while TS-type contributions mainly involve hard, nearly lightlike photons. 
In both cases, the HTL approximation remains well justified.
On the contrary, TT-type contributions significantly differ from the HTL-resummed result, since the photon propagator picks up a non-zero width when accounting for the 1PI-resummed photon self-energies, which relaxes the kinematic suppression present otherwise.
Furthermore, even for soft ALP momentum, TT-type contributions receive their dominant contributions from two hard photons---a regime where HTLs become unreliable.
As discussed in the previous section, we find that $\Pi_{TT}^< (P) \sim 1$ remains constant for a soft ALP momentum, as long as the ALP momentum $p$ still exceeds the photon width $\Gamma_{TT} \sim - \text{Im} \Pi_{T} (K) / k^0$.
For $p < \Gamma_{TT}$, we find $\Pi_{TT}^< \sim p^2$, such that TT-type contributions can constitute the dominant contribution for $\Gamma_{TT} \lesssim p \ll m_V$.
This is precisely what we observe in Fig.~\ref{fig:FullAxProd} for $p \lesssim 10^{-2} T \approx g_1^4 T$. 
For the parameters we have chosen, we find $\Gamma_{TT} \approx 5 \cdot 10^{-4} T$ below which TT-type contributions are more strongly suppressed with $p$ than SS-type contributions such that the latter eventually dominate for even smaller ALP momentum. 
In Fig.~\ref{fig:TTEstim} in the appendix, we present a plot of the evolution of the ALP self-energy for the transverse TT-type contribution including momenta $p< 10^{-3} T$.

In summary, based on our calculation\footnote{We remind the reader that our calculation captures only part of the leading-order contribution to the interaction rate for $p \lesssim g_1^2 T$. A follow-up work will address the missing contributions to provide a complete leading-order result.} we find that the production rate of ALPs from photons varies significantly across different momentum regimes, with different channels providing the dominant contribution depending on the ALP momentum:
\begin{enumerate}[label=(\roman*)]
    \item $\mathbf{p \gtrsim g_1 T}$: TS-type configurations dominate the interaction rate. 
    \item $\mathbf{g_1^2 T \lesssim p \lesssim g_1 T}$: SS-type configurations become increasingly relevant and provide a non-negligible contribution to the interaction rate.
    \item $\mathbf{g_1^4 T \lesssim p \lesssim g_1^2 T}$: SS-type contributions, well described by HTL-approximated photon propagators, primarily drive the interaction rate, as TS-type contributions become exponentially suppressed, while TT-type contributions play a noticeable role.
    \item $\mathbf{p \lesssim g_1^4 T}$: TT-type contributions, only present when considering the 1PI-resummed photon propagator, dominate the interaction rate.
\end{enumerate}
\begin{figure}[t!]
    \centering\includegraphics[width=0.7
    \textwidth]{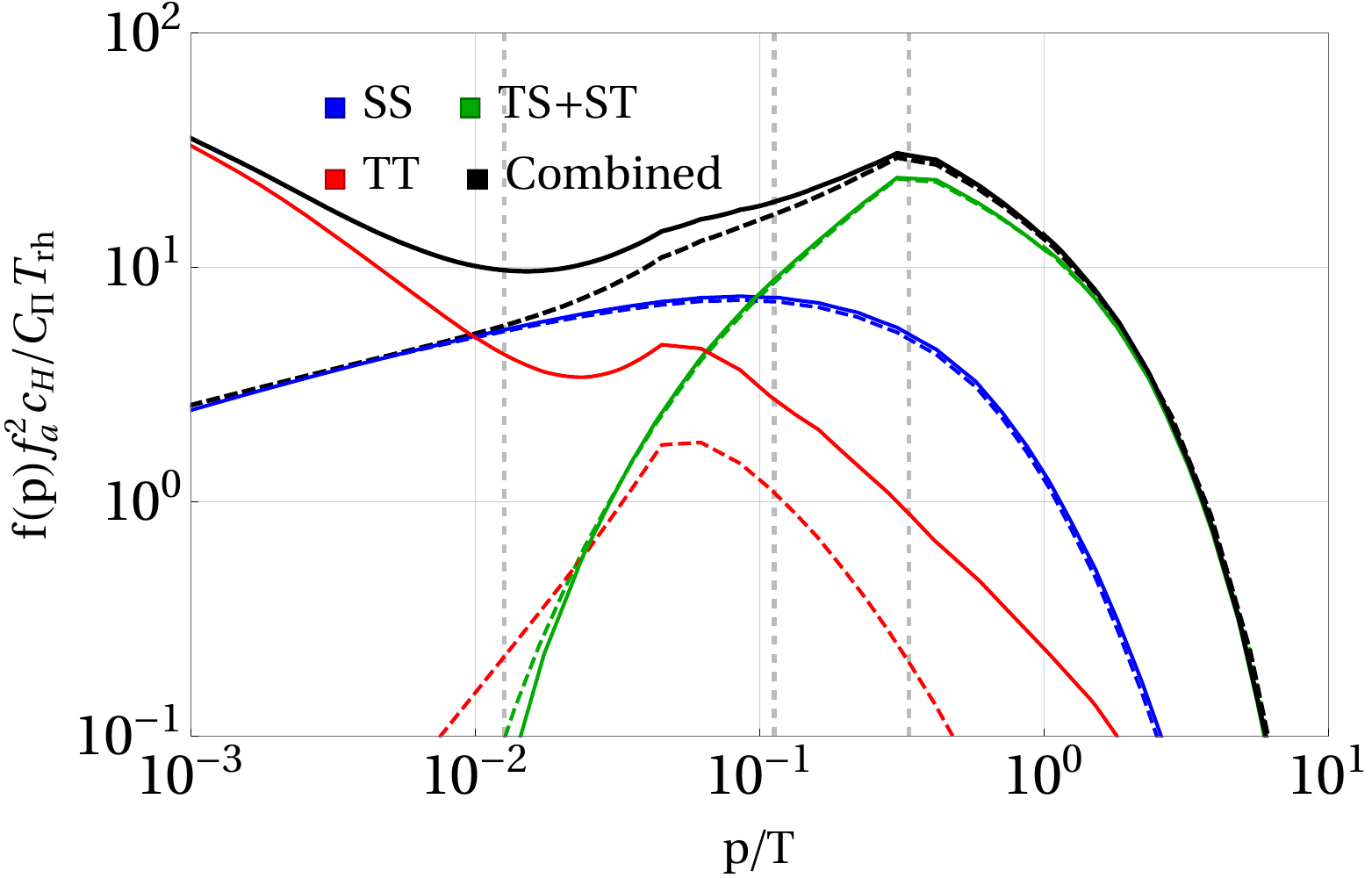}
    \caption{ALP distribution function. The dashed lines correspond to the calculation using the HTL-resummed propagator and the solid ones with the 1PI-resummed one. The dashed vertical grid lines correspond to the momentum scales $p = g_1^4 T$, $p = g_1^2 T$ and $p = g_1 T$.}
    \label{fig:AxDitribComp}
\end{figure}

\paragraph{ALP distribution function and impact on Lyman-$\alpha$.}
With the ALP interaction rate at hand, we can now solve the ALP evolution equation and construct our observables of interest.
To obtain the ALP distribution function, we integrate Eq.~\eqref{eq:BoltzTFT} over $z$:
\begin{equation}
    f(p)=\frac{m_a C_\Pi}{2 c_H f_a^2} \int_{\zrh}^{z_0} dz \frac{\Pi^{<}_a (p,z)}{z^2 \sqrt{p^2 + z^2}} \, . \label{eq:Momentum_Master_Eq}
\end{equation}
with
\begin{equation}
    c_H = \sqrt{\frac{4 \pi^3}{45}} \sqrt{g_\text{eff}} \,m_\text{Pl}^{-1} \,,
\end{equation}
where $m_\text{Pl} \approx 1.22 \times 10^{19}\,\text{GeV}$ is the Planck mass.
The integral limits are given by $z_0 = m_a / T_0$, where $T_0$ is the temperature today, and $z_\text{rh} = m_a / T_\text{rh}$, where $T_\text{rh}$ is the reheating temperature. 
From Eq.~\eqref{eq:BoltzTFT}, in combination with the results for the photon self-energy discussed above, it is evident that the production rate is dominated by $z \rightarrow 0$ contributions. 
We have only obtained numerical results for the ALP interaction rate at $z=0$ when considering the 1PI-resummed photon propagator.
While the interaction rate in the high-temperature limit is sufficient to obtain an accurate result for ultraviolet freeze-in, in our numerical implementation, we modify Eq.~\eqref{eq:Momentum_Master_Eq} by adding the expected exponential suppression for temperatures $T < m_a$, and we evaluate
\begin{equation}
    f(p)=\frac{m_a C_\Pi}{2 c_H f_a^2} \int_{\zrh}^{z_0} dz \frac{\Pi^{<}_a (p,0)}{z^2 \sqrt{p^2 + z^2}} e^{-z}\, . \label{eq:Momentum_Master_EqZ0}
\end{equation}
In Fig.~\ref{fig:AxDitribComp}, we illustrate and compare the ALP distribution function obtained using the 1PI-resummed photon propagator (solid lines) with that obtained using the HTL-resummed photon propagator (dashed lines). 
As expected, both approaches yield a positive distribution function across all momentum scales.
Furthermore, the results for hard momentum agree at the $\mc{O} (10 \, \%)$ level.
The main difference arises at soft momentum, where the result using the 1PI-resummed propagator is dominated by TT-type contributions, which are strongly suppressed in the HTL approximation. 
The latter leads to an ALP distribution function dominated by SS-type contributions at soft momentum.
Additionally, we stress that TS-type contributions, which can be identified with t-channel diagrams that in the free theory indicate the need for resummation and for which the matching procedure (as outlined in references \cite{PhysRevLett.66.2183,Bolz:2000fu,Graf:2010tv}) is typically performed, are subdominant at momenta $p \ll m_V$.

\begin{figure}[t!]
    \centering
    \includegraphics[width=0.7\textwidth]{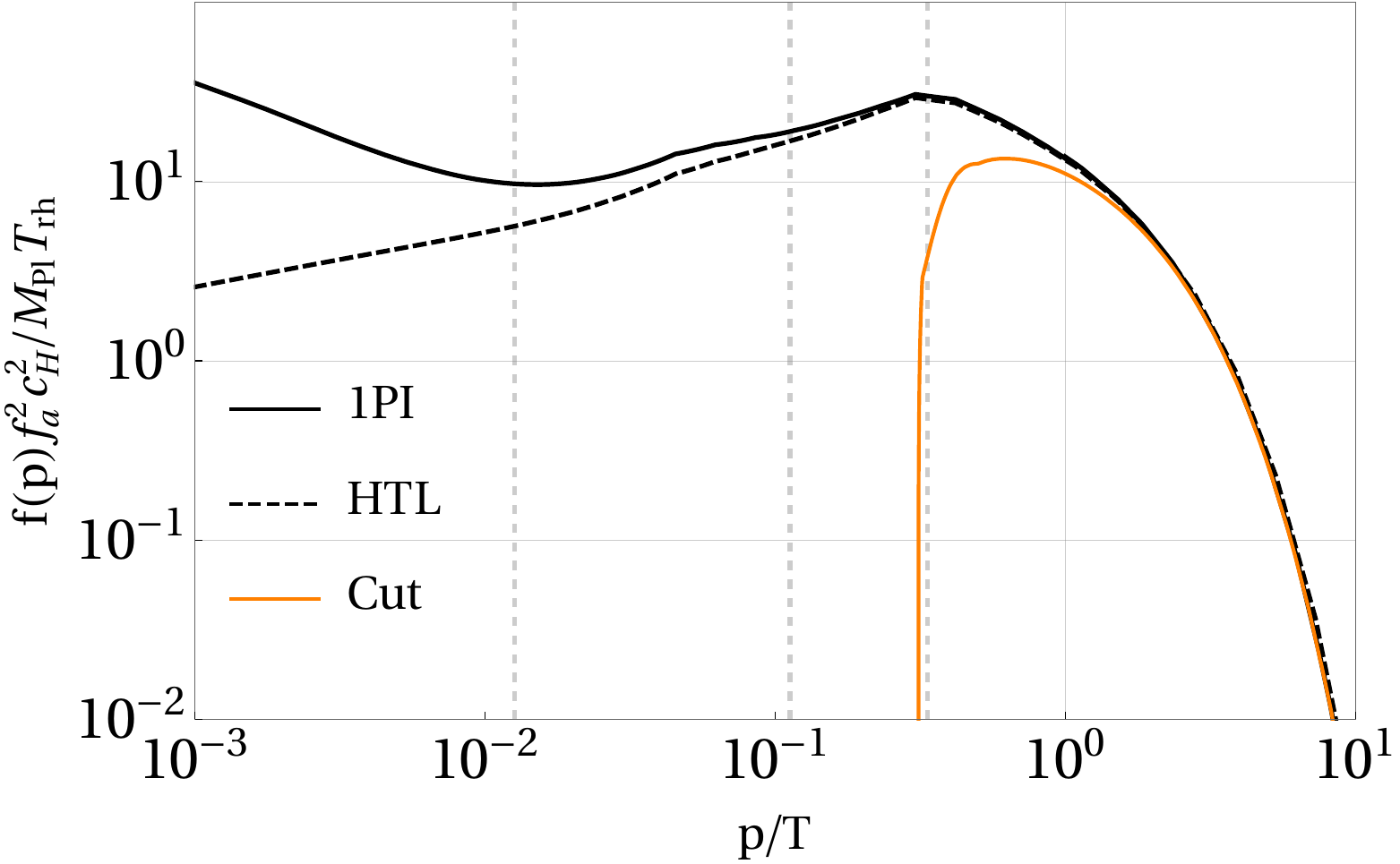}
    \caption{Comparison of the ALP distribution function obtained with the methods in this work (solid and dashed black lines) to the results of~\cite{Baumholzer:2020hvx} (orange line). The vertical dashed gray lines indicate the characteristic momentum scales $g_1 T$, $g_1^2 T$, and $g_1^4 T$.}
    \label{fig:CompPlot}
\end{figure}

We conclude this section by comparing our results for the ALP distribution functions with those from previous work~\cite{Baumholzer:2020hvx}. 
The comparison is illustrated in Fig.~\ref{fig:CompPlot}.
Our approach, using 1PI-resummed and HTL-resummed propagators (black solid and dashed lines, respectively), consistently produces positive interaction rates across all momentum scales, including the soft momentum regime ($p < g_1 T$). 
In contrast, the method based on matching HTL-resummed and free-theory propagators (orange line, labeled ``Cut'') results in unphysical negative interaction rates in this region. 
This ``Cut'' method, as used in~\cite{Baumholzer:2020hvx}, matches HTL-resummed and free propagators at an intermediate scale  $g_1 T \ll k^* \ll T$, following~\cite{Bolz:2000fu}.
For hard momenta ($p \gtrsim T$), all methods show good agreement.%
\footnote{We have corrected a missing factor of $1/2$ in the results of Ref.~\cite{Baumholzer:2020hvx}.}

To assess the impact of these differences on phenomenologically relevant quantities, we examine the number density, \( n \sim \int p^2 f_a (p) \), and the average ALP momentum, defined as
\begin{align}
    \left\langle p \right\rangle = \frac{\int dp \, p^3 f_a(p)}{\int dp \, p^2 f_a(p)} \, .
\end{align}
For the number density, we find the following ratios:
\begin{align}
    \frac{n_\text{1PI}}{n_\text{HTL}} = 0.97 \, , \quad \frac{n_\text{1PI}}{n_\text{Cut}} = 1.09 \, .
\end{align}
The average ALP momentum, calculated by numerically solving the distribution function for momenta \( p/T \in [10^{-3}, 10] \) and extrapolating beyond this range using an exponential suppression \( f(p) \sim \exp(-p/T) \), yields the following values:
\begin{align}
    \left\langle p/T \right\rangle_\text{1PI} &= 3.06 \, , & \left\langle p/T \right\rangle_\text{HTL} &= 3.17 \, , \nonumber \\
    \left\langle p/T \right\rangle_\text{Cut} &= 3.14 \, , &  \left\langle p/T \right\rangle_\text{BE} &= 2.70 \, .
\end{align}
The average ALP momentum differs by \(\mathcal{O}(1-10\%)\) across methods. 
The subscript "BE" refers to the average momentum of a particle species following a Bose-Einstein distribution and is given as in indication for the average ALP momentum in thermal equilibrium.  

It is important to study the impact of the newly obtained distribution functions on the lower bounds on the ALP mass as a warm dark matter (WDM) candidate. 
A rigorous analysis would require comparing the matter power spectrum obtained with our distribution to that of a thermal relic DM with the same mass. 
Here we adopt a simpler approach, which is valid as long as the distribution $f_a(p)$ does not deviate too much from a thermal one.
The ALP free-streaming length can be approximated as~\cite{Kolb:1990vq}
\begin{equation}
    \lambda_\mathrm{fs} \simeq 2\frac{a_0}{a_\mathrm{NR}}t_\mathrm{NR} \,,
\end{equation}
where the suffix NR indicates the time at which the ALP becomes non-relativistic, defined as $\langle p \rangle = m_a$. One thus easily sees that
\begin{equation}
    \lambda_\mathrm{fs} \propto T_\mathrm{NR}^{-1} \propto \frac{\langle p \rangle /T}{m_a} \,.
\end{equation}
Thus, a percent level decrease in $\langle p \rangle /T$ with respect to the results of Ref.~\cite{Baumholzer:2020hvx} translates into a correspondingly weaker lower bound on the mass of ALPs as WDM. 
The impact of avoiding the unphysical results at low $p$ is limited, thanks to the fact that the average is dominated by larger momenta.

\section{Conclusions}\label{sec:Conclusions}
In this article, we analyzed the production rate of an Axion-Like Particle (ALP) coupled to an abelian gauge boson with a gauge coupling $g$, placing special emphasis on soft ALPs, i.e. those with momenta $p \lesssim g T$. 
While our results are general and apply to any model of this kind, including QED or a dark photon, we specifically focused on the vector boson $B_\mu$ of $\mathrm{U}(1)_Y$ throughout this work. 

Previous studies have addressed infrared divergences arising in t-channel diagrams by resumming the problematic gauge boson propagator in the so-called Hard Thermal Loop (HTL) approximation. 
The HTL-resummed propagator provides a reliable approximation to the full form of the 1PI-resummed propagator when the gauge boson four momentum $K$ is either soft $K \lesssim g T$, or nearly lightlike $K^2 \lesssim g^2 T^2$. 
However, the production of ALPs involves gauge bosons across the entire momentum range. As a result, the HTL resummed propagator becomes unreliable when applied outside its domain of validity.
Thus, prior approaches typically use the free-theory propagator outside the HTL regime or implement a subtraction scheme, where the kinematic regions satisfying the HTL conditions are subtracted from the free-theory interaction rate, and the HTL-resummed contribution is added back.
Although this subtraction scheme is accurate for hard ALP momentum, it leads to unphysical, negative interaction rates at soft ALP momenta.

Here, we address this issue by considering the full form of the 1PI-resummed gauge boson propagator, which is valid and reliable across all momentum scales.
This approach eliminates the need for any matching or subtraction procedures to calculate the ALP interaction rate. 
Importantly, our results demonstrate that this method yields positive interaction rates for all momenta. 
This represents an improvement over methods that rely on HTLs, offering a more accurate prediction for ALP production.

Our calculation of the ALP production rate is based on the imaginary part of its retarded self-energy at the one-loop level and includes two new important kinematic configurations for soft ALPs: 
ALP production from two soft, spacelike gauge bosons (SS-type) and production from two hard, timelike gauge bosons (TT-type). 
Notably, these contributions require the resummation of both gauge boson propagators in the ALP self-energy. 
In particular, TT-type contributions only arise when the full form of the 1PI-resummed propagator is used, as the finite gauge boson width is necessary to lift the kinematic constraints that otherwise strongly suppress this contribution. 
Both the SS- and TT-type contributions are  power-law suppressed at small ALP momentum, in contrast to contributions from one timelike and one spacelike gauge boson (TS-type), which dominate at hard momentum but become exponentially suppressed at small ALP momentum.
As a result, SS-type contributions, which scale as $p^{4/3}$ with the soft ALP momentum in the ALP self-energy, become significant for $p \lesssim g T$, and dominate when $p \lesssim g^2 T$.
At even smaller momentum scales, $p \lesssim g^2 T$, TT-type contributions become increasingly important and eventually dominate the production rate for $p \lesssim g^4 T$.
Neglecting SS- and TT-type contributions for soft ALP momenta appears to underestimate the interaction rate by more than one order of magnitude.
Nevertheless, we stress that this behavior needs to be confirmed by a consistent leading order computation in this regime, which must account for the diagrams shown in the last row of Fig.~\ref{fig:Cuts_and_Diagrams_included}, as discussed at the end of Sec.~\ref{sec:AxionProduction}.   

Although these corrections are crucial for soft ALPs, their impact on phenomenological observables like number density or average momentum is limited, as soft momenta are suppressed in integrated quantities.
For instance, we observe a modest reduction (increase) in the average ALP momentum (ALP number density) when using the full 1PI-resummed propagator compared to calculations relying on free gauge boson propagators for hard momenta.

The results presented in this work offer a consistent and reliable framework for calculating ALP production rates across a wide range of momentum scales. 
While our approach does not strictly apply to ALP momenta $p < g^4 T$, where the quasiparticle description breaks down and hydrodynamics become necessary, it provides positive and physically meaningful interaction rates in the range $g^4 T \lesssim p \lesssim g T$. 
This represents an important step toward achieving full leading-order accuracy in this momentum regime, as discussed in Sec.~\ref{sec:AxionProduction}. 
Extending this work to achieve full leading-order accuracy in this regime represents an exciting challenge for future research. 
The same holds for ALPs coupling to non-Abelian gauge bosons, where two-loop vertex corrections to the ALP self-energy, no longer suppressed by the feeble ALP coupling, must be incorporated.


\acknowledgments
We would like to thank E.~Copello for collaboration in initial stages of this project. We thank J.~Ghiglieri, M.~Laine, D.~Boedeker, M.~Fernandez Lozano for very helpful discussions. 
We acknowledge support by the Cluster of Excellence “Precision Physics, Fundamental Interactions, and
Structure of Matter” (PRISMA$^+$ EXC 2118/1) funded by the Deutsche Forschungsgemeinschaft (DFG, German Research
Foundation) within the German Excellence Strategy (Project No. 390831469).
M.~B. and J.~H. acknowledge support from the Deutsche Forschungsgemeinschaft (DFG, German Research
Foundation) Emmy Noether grant HA 8555/1-1.
We are grateful to the Mainz Institute for Theoretical Physics (MITP) of the Cluster of Excellence PRISMA+, and to the organizers of the workshop ``The Dark Matter Landscape: From Feeble to Strong Interactions'', for the hospitality and the support during the completion of this work.
Furthermore, this work is supported in part by the Italian MUR Department of Excellence grant 2023-2027 “Quantum Frontiers”.
E.~M. acknowledges support from the Italian Ministry for University and Research (MUR) Rita Levi-Montalcini grant ``New directions in axion cosmology''.
M.~B. and E.~M. are supported by Istituto Nazionale di Fisica Nucleare (INFN) through the Theoretical Astroparticle Physics (TAsP) project.


\appendix
\setcounter{equation}{0}
\section{Abelian Gauge Boson Self-Energy}\label{app:A}

In this section, we  provide a detailed presentation of the procedure and expressions required for calculating the ALP self-energy, employing the complete form of the resummed Wightmann propagator in Eq.~\eqref{eq:non-time-propagator.}. \\

We decompose the propagator into different polarization components: transverse, longitudinal, and parallel to $K=(k_0,\Vec{k)}$, i.e the gauge boson four-momentum. The Lorentz structures of the corresponding spectral propagators are: 
\begin{subequations}
    \begin{align}
    \mathcal{P}_{\mu\nu}^t&= -\tilde{\eta}_{\mu\nu}+\frac{\tilde{K}_{\mu\nu}\tilde{K}_{\nu}}{-k^2}\\
    \mathcal{P}_{\mu\nu}^l&= -\eta_{\mu\nu}+\frac{K_{\mu\nu}K_{\nu}}{-K^2}-    \mathcal{P}_{\mu\nu}^t\\
    \mathcal{P}_{\mu\nu}^K&=-\frac{K_\mu K_\nu}{K^2}
    \end{align}
\end{subequations}
where $\tilde{\eta}_{\mu\nu}=\eta_{\mu\nu}-U_\mu U_\nu$, and $\tilde{K}_\mu=K_\mu -(K \cdot U)U_\mu$. 
The vector $U$ represents the  four-velocity of the plasma.\\

As stated in the main text, when we do the full calculation we need to add the vacuum photon self-energy $\pi_0$ to ensure that the spectral density has the proper sign, $\text{sign}(\rho_{l/t}(K))=\text{sign}(k_0)$.
The photon self-energy in vacuum results in
\begin{align}\label{eq:ImVaacSE}
    \Im\pi_0 &=\frac{g^2K^2}{48\pi^2}(2N_F+N_S)\Theta(K^2) \, , & 
   \Re\pi_0 &=\frac{g^2K^2}{48\pi^2}(2N_F+N_S)\log\left(\frac{-K^2}{\mu}\right) \, .
\end{align}
We only include the imaginary part of it in our numerical results calculations since we verified that the real part is negligible when the renormalization scale is chosen as $\mu=T$.\\
 
It is as well necessary to use the integrals $H_S$, $H_F$, $G_S$ and $G_F$ appearing in the Eqs.~\eqref{eq:photonSE} in their non-approximate form. 
Specifically, we are interested in their real and imaginary parts. 
The latter is analytical and yields
{\scriptsize
\begin{subequations}
\begin{align}\label{ImHF}
     \Im H_F &=  \int_0^\infty\frac{dp}{2 \pi^2}f_{FD}(p)\left[2p\frac{k_0}{k}\pi\Theta(-K^2)+\frac{\pi}{k} (p-k_0^+)(p+k_0^-)\Theta(p-k_0^-)\Theta(-p+k_0^+)\right .\nonumber \\
      & \left .-\frac{\pi}{k}(p+k_0^+)(p+k_0^-)\Theta(p+k_0^+)\Theta(-p-k_0^-)\right]\\
      &= \theta(K^2)\left[\frac{k(Li_2[-e^{-k_0^-}]+Li_2[-e^{-k_0^+}])+2(-Li_3[-e^{-k_0^-}]+Li_3[-e^{-k_0^+}])]}{2k\pi}\Theta(k_0)\right . \nonumber\\
      &\left .\frac{k(Li_2[-e^{k_0^-}]+Li_2[-e^{k_0^+}])+2(Li_3[-e^{k_0^-}]-Li_3[-e^{k_0^+}])]}{2k\pi}\Theta(-k_0) \right] \nonumber\\
      &+\Theta(-K^2)\frac{k(Li_2[-e^{-k_0^+}]-Li_2[-e^{k_0^-}])+2(Li_3[-e^{-k_0^+}]+Li_3[-e^{k_0^-}])]}{2k\pi},
\end{align}
\end{subequations}

\begin{subequations}
\begin{align}\label{ImHS}
     \Im H_S &=  \int_0^\infty\frac{dp}{2 \pi^2}f_{B}(p)\left[2p\frac{k_0}{k}\pi\Theta(-K^2)+\frac{\pi}{k} (p-k_0^+)(p+k_0^-)\Theta(p-k_0^-)\Theta(-p+k_0^+)\right .\nonumber \\
      & \left .-\frac{\pi}{k}(p+k_0^+)(p+k_0^-)\Theta(p+k_0^+)\Theta(-p-k_0^-)\right .\nonumber \\
      & \left . -\frac{k}{4}\pi(\Theta(p+k_0^+)\Theta(-p-k_0^-)+\Theta(p-k_0^-)\Theta(-p+k_0^+))\right]\\
      &=\theta(K^2)\left[-\frac{k^2\left(k+\log\frac{-1+e^{k_0^-}}{-1+e^{k_0^+}}\right)+4k(Li_2[e^{-k_0^+}]+Li_2[e^{-k_0^-}])+8(-Li_3[e^{-k_0^+}]+Li_3[e^{-k_0^-}])]}{8k\pi}\Theta(k_0)\right . \nonumber\\
      &\left .\frac{k^2\left(\log\left[\frac{1-e^{k_0^+}}{1-e^{k_0^-}}\right]\right)+4k(Li_2[e^{k_0^-}]+Li_2[e^{k_0^+}])+8(Li_3[e^{k_0^-}]-Li_3[e^{k_0^+}])]}{8k\pi}\Theta(-k_0) \right] \nonumber\\
      &-\Theta(-K^2)\frac{k^2\left( k+w+\log\frac{1-e^{k_0^-}}{-1+e^{k_0^+}}\right)+8k(Li_2[e^{-k_0^+}]-Li_2[e^{k_0^-}])+16(Li_3[e^{-k_0^+}]+Li_3[e^{k_0^-}])]}{16k\pi},
\end{align}
\end{subequations}
\begin{subequations}
\begin{align}\label{ImGF}
     \Im G_F &=\int_0^\infty\frac{dp}{2 \pi^2}f_{FD}(p)\left[4p-\pi\frac{K^2}{2k}(\Theta(p+k_0^+)\Theta(-p-k_0^-)-\Theta(p-k_0^-)\Theta(-p+k_0^+))\right]\\
      &= \Theta(K^2)\left[-\frac{(k-k_0)(k+k_0)(k+\log[1+e^{k_0^-}]-\log[1+e^{k_0^+}]}{4k\pi}\Theta(k_0)\right .\nonumber \\
      & \left .+\frac{(k-k_0)(k+k_0)(k+\log[1+e^{k_0^+}]-\log[e^k+e^{k_0^+}]}{4k\pi}\Theta(-k_0)\right]\nonumber \\
      &-\Theta(-K^2)\frac{(k-k_0)(k+k_0)(k+k_0+2(\log[1+e^{k_0^-}]-\log[e^k+e^{k_0^+}])}{8k\pi},
\end{align}
\end{subequations}
\begin{subequations}
\begin{align}\label{ImGS}
     \Im G_S &=\int_0^\infty\frac{dp}{2 \pi^2}f_{B}(p)\left[4p+\pi\frac{K^2}{4k}(\Theta(p+k_0^+)\Theta(-p-k_0^-)-\Theta(p-k_0^-)\Theta(-p+k_0^+))\right]\\
      &= \Theta(K^2)\left[-\frac{(k-k_0)(k+k_0)(k+\log[-1+e^{k_0^-}]-\log[1-e^{k_0^+}]}{8k\pi}\Theta(k_0)\right .\nonumber \\
      & \left .+\frac{(k-k_0)(k+k_0)(-\log[1-e^{k_0^-}]+\log[1-e^{k_0^+}]}{8k\pi}\Theta(-k_0)\right]\nonumber \\
      &-\Theta(-K^2)\frac{(k-k_0)(k+k_0)(k+k_0+2(\log[1-e^{k_0^-}]-\log[-1+e^{k_0^+}])}{16k\pi}.
\end{align}
\end{subequations}
}
In these equations, we have used $k_0^+=(k_0+k)/2$ and $k_0^-=(k_0-k)/2$, and
\begin{equation}
    f_{FD}=\frac{1}{1+e^{k_0}} \, ,
\end{equation}
is the Fermi-Dirac distribution.

The real parts of the integrals are more complicated to evaluate since they have a part that only can be calculated numerically. For that we integrate the analytical part and identify the numerical parts:

\begin{align}
   \label{ReHF} \Re (H_F) &=  \frac{1}{12} \left( 1 - \frac{k_0}{k} \log \left| \frac{k_0^+}{k_0^-} \right| \right) + \frac{1}{2 \pi^2 k} I_{3+} ( k_0, k ) \, ,\\
   \label{ReHS}  \Re (H_S) &= \frac{1}{6} \left( 1 - \frac{k_0}{k} \log \left| \frac{k_0^+}{k_0^-} \right| \right) + \frac{1}{2 \pi^2 k} I_{3-} ( k_0, k ) + \frac{k}{8 \pi^2} I_{2-} (k_0, k) \, , \\
    \label{ReGF} \Re (G_F) &= \frac{1}{6} + \frac{K^2}{4 \pi^2 k} I_{2+} (k_0, k) \, , \\
    \label{ReGS}\Re(G_S) &= \frac{1}{3} - \frac{K^2}{8 \pi^2 k} I_{2-} (k_0, k) \, ,  
\end{align}
with
\begin{align}
    I_{2\pm} (k_0, k) &= a_\pm (k_0^+) - a_\pm (k_0^-) \, , \\
    I_{3\pm} (k_0, k) &= k_0^+ k_0^- \left[ a_\pm (k_0^+) - a_\pm (k_0^-) \right] + \left[ b_\pm (k_0^+) - b_\pm (k_0^-) \right] \nonumber\\
    &+ \left( k_0^+ + k_0^- \right) \left[ c_\pm(k_0^+) - c_\pm(k_0^-) \right] \, ,
\end{align}
and
\begin{align}
   a_\pm (y) &= \int \limits_0^\infty \dd p  \, \log \left| \frac{p+y}{p-y} \right| f_{B/FD}(p) \, , \\
   b_\pm (y) &= \int \limits_0^\infty \dd p \, p^2 \log \left| \frac{p+y}{p-y} \right| f_{B/FD} (p) \, , \\
   c_\pm (y) &= \int \limits_0^\infty \dd p \, p \log \left| p^2 -y^2 \right| f_{B/FD} (p) \, .
\end{align}
We tabulate the values of the integrals $a_\pm$, $b_\pm$, and $c_\pm$ with $2000$ ($10000$ for $c_\pm$) logarithmic steps between $y \in [10^{-6},10^{4.5}]$, using analytic approximations outside of this interval.\\
Once we have these integrals we can evaluate the spectral densities in the Eqs.~\eqref{eq:spectraldensT} and \eqref{eq:spectraldensL} and numerically solve the ALP self-energy.


\section{Approximate Solutions to the ALP Self-Energy} \label{app:B}
In this appendix, we provide a brief derivation of the estimates for the scaling behavior of different kinematic configurations in the soft ALP momentum limit, \( p \ll m_V \).
\subsection{Approximate Solutions: Double-Spacelike}
The double-spacelike contribution is given by Eq.~\eqref{eq:TL-axionSE}, evaluated at spacelike $K$ and $Q$.  
Here, we estimate the scaling behavior in the $p^0 = p \ll m_V$ limit of the dominant contribution involving two transverse gauge bosons.  
We find that, as long as $k^0, k \lesssim m_V$\footnote{$k^0, k \gtrsim m_V$ contributions are power-law suppressed at the level of the spectral propagator and cannot be Bose-enhanced since $p \ll m_V$.},  
our integral is dominated by configurations with $k^0 \ll k$. In this limit, the spectral propagator is well approximated by HTLs and takes the form  
\begin{align}
    \rho_T (K) &= \pi m_V^2 \frac{k^0}{k^5} \left[ 1 - \left( \frac{k^0 }{k} \right)^2 \frac{ m_V^4 \pi^2 + 16 m_V^2 k^2 - 4 k^4}{4 k^4} + \mathcal{O} \left( \left( \frac{k^0}{k} \right)^4 \right)\right] \nonumber \\
    &\approx \pi m_V^2 \frac{k^0}{k^5} \left[ 1 - \left( \frac{k^0 }{k} \right)^2 \frac{ m_V^4 \pi^2}{4 k^4} + \mathcal{O} \left( \left( \frac{k^0}{k} \right)^4 \right)\right] \, , \label{eq:spectralSSapprox}
\end{align}
where we assumed $k \ll m_V$ in the last line.  
For fixed $k^0$, the spectral propagator is then maximized (again, expanding in $k^0 \ll k$) for  
\begin{align}
    k_\text{max} = \left( \frac{11 \pi^2 m_V^4}{20} \right)^\frac{1}{6} \left( k^0 \right)^\frac{1}{3} \, .
\end{align}
Thus, for the estimate of the double-spacelike contribution we assume the integral to be dominated by momenta $k \sim \mathcal{O} \left( k_\text{max} \right)$. 
Thus, for the estimate of the double-spacelike contribution, we assume the integral to be dominated by momenta $k \sim \mathcal{O} \left( k_\text{max} \right)$.  
The distribution functions appearing in the ALP self-energy can be Bose-enhanced as long as $k^0 \lesssim \mathcal{O} (p)$, and thus, we expect the dominant contribution to arise from those energies.  

Additionally, the integration over the momentum $q$ is restricted to the domain $\left| p - k \right| \leq q \leq \left| p + k \right|$, and since $k_\text{max} \gg p$, we can approximate the integral over $q$ as the width of the integration interval times the integrand evaluated at $q = k$.  
This implies that the second spectral propagator, $\rho_T(Q)$, can also be evaluated using Eq.~\eqref{eq:spectralSSapprox}, since $q^0 = p^0 - k^0 \sim \mathcal{O} (p) \ll q$.  
Finally, the square bracket in Eq.~\eqref{eq:LT-TTcontrib} must be expanded in $p \ll k$.  
Eventually, we find  
\begin{align}
    \Pi_{a,SS}^< (p \ll m_V) &\approx \frac{2 p c_1 \alpha_1}{f_a^2  8 (2 \pi)^5 p} \int \limits_{\mc{O}(p)} dk^0 \int \limits_{\mc{O}(k_\text{max})} dk \, \pi^2 m_V^4 \frac{k^0}{k^5} \frac{p^0 - k^0}{k^5} \frac{8 k^4 p^2 T^2}{k^0 \left( p^0 - k^0 \right)} \nonumber \\
    &\approx \kappa_I \frac{c_1 \alpha_1}{f_a^2  8 (2 \pi)^5 }  \frac{24}{5} \left( \frac{20}{11} \right)^\frac{5}{6} \pi^\frac{1}{3} T^2 m_V^\frac{2}{3} p^\frac{4}{3}  \, ,
\end{align}
where $\kappa_I$ is a (constant) numerical factor parameterizing the inaccuracy of our estimate.  
We find good numerical agreement when choosing $\kappa_I = 8$, resulting in  
\begin{align}
    \Pi_{a,SS}^< (p \ll m_V) &\approx 93 \frac{c_1 \alpha_1}{f_a^2 8 (2 \pi)^5} T^2 m_V^{\frac{2}{3}} p^{\frac{4}{3}} \, .
\end{align}
We can observe this behavior by looking at the blue lines in Fig.~\ref{fig:Est}.
\subsection{Approximate Solutions: Double-Timelike}
Here, we give the soft momentum limit $p = p^0 \ll m_V$ of the ALP self-energy in Eq.~\eqref{eq:TL-axionSE} for $K^2, Q^2 > 0$ and two transverse gauge bosons, which constitute the dominant contribution to the interaction rate.  
For a lightlike ALP momentum $P$, this contribution is only sizable if the full form of the gauge boson self-energy is considered, since the imaginary part of the self-energy vanishes in the HTL approximation.  
As the full form of the self-energy has a complicated analytic structure, or in the case of the real part is only given in terms of a one-dimensional integral, a purely analytic estimate of this quantity is hardly possible.  
Thus, we rely on an approximate method (see for instance section 4.2 of \cite{Drewes:2013iaa}) that extends the $k^0$-integration in Eq.~\eqref{eq:TL-axionSE} into the complex plane.  
Then, using the residue theorem, we find the result of the $k^0$ integration by extracting the complex poles in $k^0$ of the two spectral propagators $\rho_T (K)$ and $\rho_T(Q)$, which in the limit of a narrow width  
\begin{align}
    \Gamma_{TT} = -\frac{\text{Im} \Pi_T (K)}{k^0} \ll k^0 \, ,
\end{align}
are approximately given by the solution $k^0_\text{pole} (k)$ to $K^2 = \text{Re}\, \Pi_T (K)$ and must be found numerically.\footnote{For sufficiently large momentum $k \gg g\, T$, the solution is simply given by the dispersion relation $k^0_\text{pole} = \sqrt{k^2 + m_V^2}$.}  
Effectively, to leading order in a vanishing width $\Gamma_{TT}$, we then find  
\begin{align}
    \Pi_{a,TT}^< (p) = 2 \frac{C_{\Pi}}{f_a^2 p} \int_0^\infty dk \int_{|p-k|}^{|p+k|} dq \, \sum_i \pi \frac{\rho_T (P - K_i)}{k^0_i} f_{TT} \left( k^0_i, k, q^0, q, P \right) \, ,
\end{align}
with  
\begin{align}
   f_{TT} \left( k^0, k, q^0, q, P \right) &= f_B (k^0) f_B(p^0 - k^0) k q \nonumber \\ 
   &\times \left[ \left(\frac{k^0}{k}\right)^2 + \left(\frac{q^0}{q}\right)^2 ((k^2- p^2 + q^2) + 4k^2q^2) + 8 k^0 q^0 (k^2 +q^2 -p^2) \right] \, ,
\end{align}
and $k^0_i \in \left \lbrace k^0_\text{pole}, -k^0_\text{pole}  \right \rbrace$.  
We find that the integral is dominated by momenta $k \gtrsim g \,T \gg p$, which again allows us to approximate the $q$-integration as  
\begin{align}
    \Pi_{a,TT}^< (p \ll g T) =4 \pi p \frac{C_{\Pi}}{f_a^2 p} \int_0^\infty dk \, \sum_\text{i}  \frac{\rho_T (P - K_\text{i})}{k^0_\text{i}} f_{TT} \left( k^0_\text{i}, k, -k^0 + p, k \pm p, P \right) \, ,
\end{align}
We expand in small $p \ll k$ and find  
\begin{align}
    f_{TT} \left( k^0_\text{pole}, k, -k^0 + p, k \pm p, P \right) \approx 8 k^4 p^2 f_B \left( k^0_\text{pole} \right) f_B\left( -k^0_\text{pole} \right) \, . 
\end{align}
The scaling behavior of $\rho_T (P - K_i)$ crucially depends on the size of $\frac{p}{\Gamma_{TT} (k^0_\text{pole})}$ since it determines if the spectral propagator is evaluated near its peak.  
If $p \ll \Gamma_{TT} (K_\text{pole})$, we can approximate  
\begin{align}
    \rho_T \left( P - K_i \right) \approx -\frac{2}{\text{Im} \Pi_T (-K_i)} \, ,
\end{align}
which is independent of $p$.  
Conversely, for $p \gtrsim \Gamma_{TT} (K_\text{pole})$  
\begin{align}
    \rho_T \left( P - K_i \right) \approx -\frac{2 \text{Im} \Pi_T (P-K_i)}{ \left[ \left( P - K_i \right)^2 - \text{Re} \Pi_T (P-K_i) \right]^2} \approx - \frac{\text{Im} \Pi_T (-K_i)}{2 p^2 \left( k^0_i - k \right)^2} \, ,
\end{align}
we find $\rho_T \sim p^{-2}$.  
Since the width obeys $\Gamma_{TT} (K_\text{pole}) \leq \Gamma_{TT} (k = 0)$, we can estimate the ALP momentum from which we can safely apply $\rho_T (P - K_i) \sim p^{-2}$, as $p_c = \Gamma_{TT} (k = 0)$.  
Overall, this implies  
\begin{align}
    \Pi_{a,TT}^< (p) \approx \frac{32 \pi C_{\Pi}}{f_a^2} \int_0^\infty dk \, \sum_\text{i}  k^4 f_B \left( k^0_i \right) f_B \left( - k^0_i \right) \times \begin{cases}
         -\frac{\Gamma_{TT} (-K_i)}{2 \left( k^0_i - k \right)^2}  & p \gtrsim p_c \\
        -\frac{2}{\Gamma_{TT} (-K_i) \left( k^0_i \right)^2} \, p^2 & p \lesssim p_c
    \end{cases} \, .
\end{align}
For the choice of $g_Y=0.35$, $N_F = 10$ and $N_S = 1/2$, we find $p_c \approx 5 \cdot 10^{-4} T$.  
Since the minimal momentum presented in our numerical analysis is $p_\text{min} = 10^{-3} T$, we expect the ALP self-energy to be momentum-independent for sufficiently small momenta.  
When extending our analysis to smaller momentum $p<10^{-3} T$, we found the predicted $p^{2}$ scaling of the ALP self-energy, as can be observed in Fig.~\ref{fig:TTEstim}.  

The integral in the expression above must be solved numerically.  
Nevertheless, we have identified the $p$-dependence of the ALP self-energy and therefore of the production rate in this kinematic regime.  

\subsection{Approximate Solution: One Timelike, One Spacelike}
\begin{figure}[h!]
  \centering
  \subfloat[Scaling Estimates]{\includegraphics[width=0.5\textwidth]{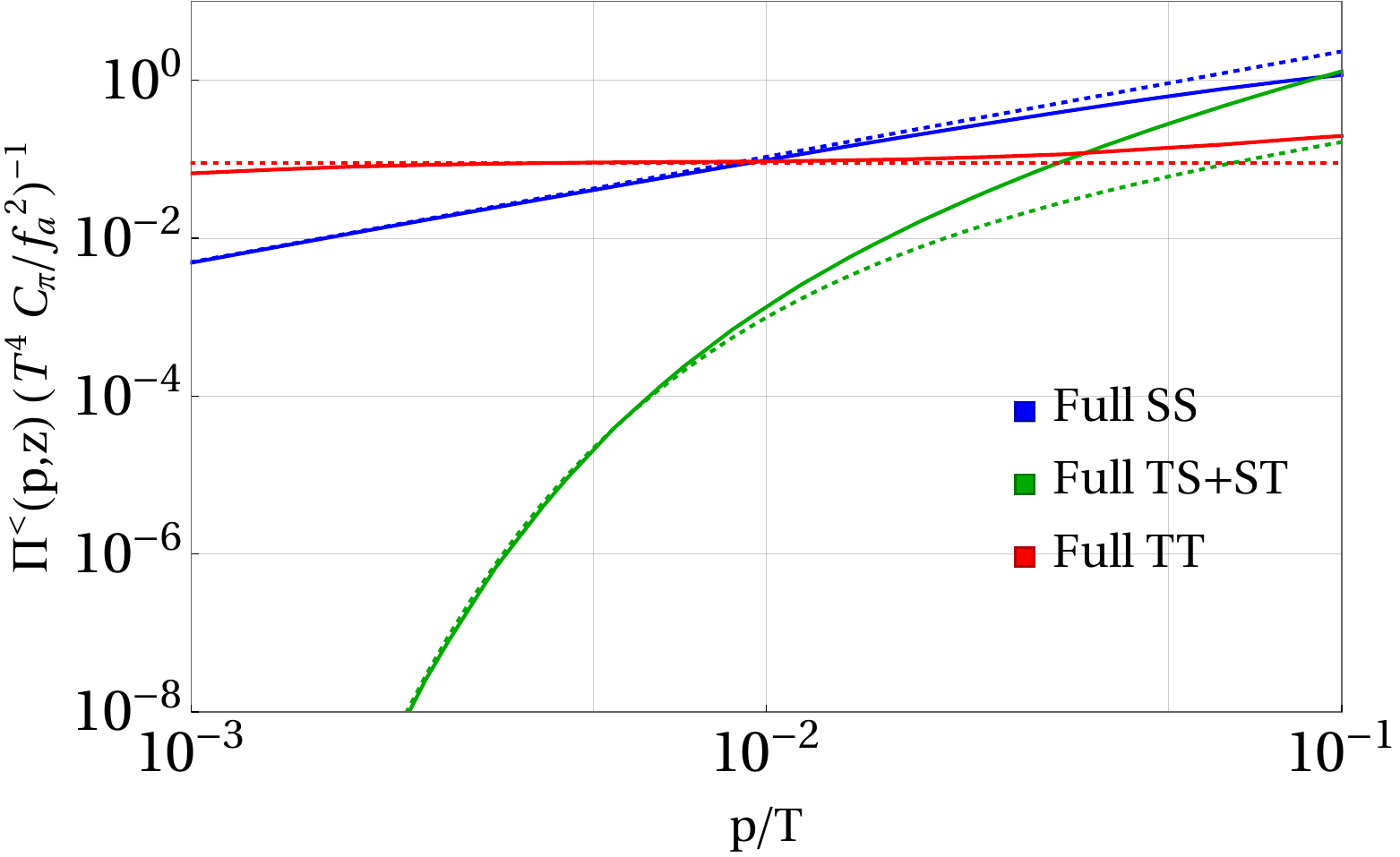}\label{fig:Est}}
  \hfill
  \subfloat[Extended Momentum Range]{\includegraphics[width=0.5\textwidth]{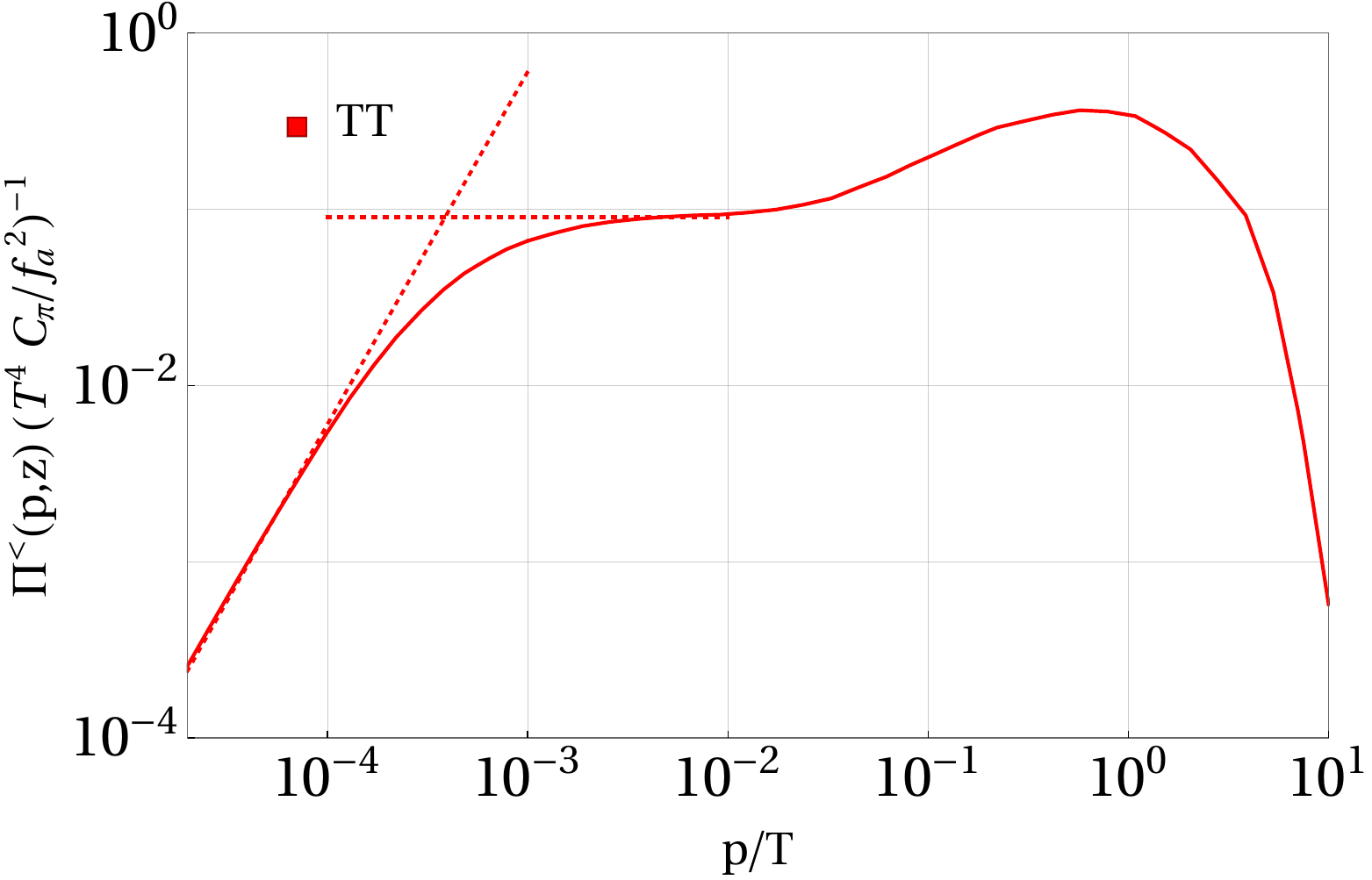}\label{fig:TTEstim}}
  \caption{Scaling estimates for the ALP production rate $\Pi^<_a(P)$ from two transverse gauge bosons (dotted lines) are compared to the full numerical counterpart (solid lines). In the left panel, we show the three different kinematic regimes SS, TS and TT. In the right panel we only show the TT regime but extending the computation until $p/T=10^{-5}$, such that the change in the scaling behavior at $p=\Gamma_{TT} \approx 5 \cdot 10^{-4}$ becomes visible. The estimates are represented by dotted lines.}
  \label{fig:Estim.}
\end{figure}

Here we identify the scaling of the ALP interaction rate from one timelike and one spacelike gauge boson with a soft momentum $p \lesssim m_V$.  
As for the double-spacelike contribution, we carry out the estimate using HTL resummed propagators and considering two transverse gauge bosons.  
For timelike momentum $K$, the HTL resummed spectral propagator takes the form of a $\delta$-function and imposes $k^0 = \pm k^0_\text{pole} (k)$, which is a solution to $K^2 = \text{Re}\,\Pi_T (K)$.  
A spacelike $Q = P - K$ with a lightlike $P \ll m_V$ requires hard momenta
\begin{align}
    k \gtrsim \frac{m_V^2}{4 p} \, , 
\end{align}
such that both momenta $K$ and $Q$ are almost lightlike, justifying the use of HTL resummed propagators.  
In this limit, again approximating the integral over $q$ by evaluating the integrand at $q=k$ and multiplying by $p$, the ALP self-energy results in
\begin{align}
    \Pi_{a,TS}^< (p) \approx  p \frac{C_{\Pi}}{f_a^2 p} \int_{\frac{m_V^2}{p}}^\infty dk \, \sum_i \pi \frac{\rho_T (P - K_i)}{k} f_{TS} \left( k^0_i, k, q^0, q, P \right) \, .
\end{align}
The spectral function can be approximated as
\begin{align}
    \rho_T (P-K_i) \approx - \frac{\pi}{4} \frac{1}{p} \frac{m_V^2}{k^3} \, ,
\end{align}
and 
\begin{align}
    f_{TS} \left( k^0_i, k, q^0, q, P \right)  \approx 32 f_B (k) f_B (-k) k^4 p^2 \, ,
\end{align}
such that
\begin{align}
    \Pi_{a,TS}^< (p) &\approx  - \frac{C_{\Pi}}{f_a^2 p} 16 \pi^2 m_V^2 p^2  \int_{\frac{m_V^2}{p}}^\infty dk \, f_B (k) f_B (-k) \\
    &\approx  \frac{C_{\Pi}}{f_a^2} 16 \pi^2 T  m_V^2 p  \exp \left( - \frac{m_V^2}{4p T} \right) \, .
\end{align}
Clearly, this kinematic regime is subdominant at momenta $p \lesssim g^2 T$. We can observe this exponential suppression in Fig.~\ref{fig:Est}.

\clearpage
\bibliographystyle{JHEP}
\bibliography{biblio}
\end{document}